\newif\ifarxiv
\arxivtrue
\newif\ifdoublecolumn
\doublecolumntrue
\ifarxiv\else\doublecolumnfalse\fi

\ifarxiv{}
  \ifdoublecolumn{}
    \documentclass[a4paper,10pt,twocolumn]{article}
    \usepackage[cm]{fullpage}
  \else{}
    \documentclass[a4paper,10pt]{article}
    \usepackage{a4wide}
  \fi{}
\else{}
  \documentclass[runningheads,pagebackrefs]{llncs}
\fi{}
\usepackage[utf8]{inputenc}
\usepackage[T1]{fontenc}
\usepackage{graphicx}

\usepackage{subcaption}

\usepackage{diagbox}

\ifarxiv{}
  \usepackage{amssymb,amsmath,amsthm}
  \newtheorem{lemma}{Lemma}
  \newtheorem{theorem}{Theorem}
  \newtheorem{observation}{Observation}
  \theoremstyle{definition}
  \newtheorem{definition}{Definition}
  \theoremstyle{remark}
  \newtheorem{example}{Example}
\else{}
  \usepackage{amssymb,amsmath} %
  \newtheorem{observation}{Observation}
\fi{}
\usepackage[numbers,sort&compress]{natbib}
\makeatletter
\def\NAT@spacechar{~}
\makeatother

\usepackage[colorlinks=true, allcolors=blue,pagebackref]{hyperref}
\usepackage[sort&compress,nameinlink,capitalize]{cleveref}

\usepackage{paralist}
\usepackage{dsfont}
\usepackage{booktabs}

\usepackage{tikz}
\usetikzlibrary{arrows}
\usetikzlibrary{shapes}
\usetikzlibrary{decorations.pathreplacing}
\usetikzlibrary{calc}
\usetikzlibrary{positioning}

\crefname{observation}{Observation}{Observations}
\crefname{figure}{Figure}{Figures}
\Crefname{figure}{Fig.}{Figs.}

\usepackage{xargs} 
\newcommandx{\set}[2][1=1]{\ensuremath{\{#1,\ldots,#2\}}}
\newcommandx{\tlog}[3][1=,3=]{\log_{#1}^{#3}(#2)}

\newcommand{\bigO}{\mathcal{O}}

\newcommand{\noSty}[1]{\ensuremath{\mathbb{#1}}}
\newcommand{\N}{\noSty{N}}
\newcommand{\Q}{\noSty{Q}}

\newcommand{\classSty}[1]{\ensuremath{\text{#1}}}
\newcommand{\NP}{\classSty{NP}}
\newcommand{\classP}{\classSty{P}}
\newcommand{\FPT}{\classSty{FPT}}
\newcommand{\fpt}{fixed-parameter tractable}
\newcommand{\fpty}{fixed-parameter tractability}
\newcommand{\W}[1]{\classSty{W}[#1]}
\newcommand{\XP}{\classSty{XP}}

\newcommand{\MSO}{\textsc{MSO}}

\newcommand{\paramSty}[1]{\ensuremath{\mathrm{#1}}}
\newcommand{\tw}{\paramSty{tw}}
\newcommand{\utw}{\paramSty{tw_\downarrow}}
\newcommand{\ltw}{\paramSty{tw_\infty}}
\newcommand{\ttw}{\paramSty{ttw}}

\newcommand{\prob}[1]{\textsc{#1}}

\newcommand{\ug}[1]{\ensuremath{#1_{\downarrow}}}
\newcommand{\TG}{\ensuremath{\mathcal{G}}}
\newcommand{\TE}{\ensuremath{\mathcal{E}}}

\newcommand{\bbT}{\ensuremath{\mathbb{T}}}
\newcommand{\width}{\ensuremath{\operatorname{width}}}
\newcommand{\abs}[1]{\ensuremath{\left|#1\right|}}

\newcommand{\tdc}{\ensuremath{\mathrm{tdc}}}

\newcommand{\eps}{\ensuremath{\varepsilon}}
\newcommand{\decprob}[3]{
	\begin{center}
		\begin{minipage}{0.96\linewidth}%
			\noindent
			\textsc{#1}
			\begin{compactdesc}
			 \item[\textbf{Input:}]  #2
			 \item[\textbf{Question:}]  #3
			\end{compactdesc}
		\end{minipage}
	\end{center}
}

\newcommand{\tps}{treewidth-preserving structure}

\newcommand{\xref}[2]{\cref{#1}\eqref{#2}}

\newcommand{\casablanca}[1]{>>\textsl{#1}<<}

\begin{document}

\title{As Time Goes By:\\Reflections on Treewidth for Temporal Graphs\thanks{Dedicated to Hans L.\ Bodlaender on the occasion of his 60th birthday. \protect\\ The inclined reader, besides hopefully discovering interesting science, is
also invited to enjoy a few quotes from a famous movie 
scattered around our text; the paper title is partially taken from the 
theme song of this movie.}}

\ifarxiv{}
\author{Till Fluschnik
\and
Hendrik~Molter
\and
Rolf~Niedermeier
\and
Malte~Renken
\and
Philipp~Zschoche
}
\date{Technische Universit\"at Berlin, Faculty~IV, Algorithmics and Computational~Complexity, Berlin, Germany\\
\texttt{\small{\{till.fluschnik,h.molter,rolf.niedermeier,m.renken,zschoche\}@tu-berlin.de}}}

\fi{}
\maketitle              %

\begin{abstract}
Treewidth is arguably the most important structural graph parameter
leading to algorithmically beneficial graph decompositions.
Triggered by a strongly growing interest in 
temporal networks (graphs where edge sets change over time),
we discuss
fresh algorithmic views
	on \emph{temporal} tree decompositions and \emph{temporal} treewidth.
We review and explain some of the recent work together with some encountered pitfalls,
and  we point out challenges for future research.
\smallskip

\end{abstract}

\section{Introduction}\label{sec:intro}

\casablanca{You must remember this:}
treewidth is one of the most important structural graph parameters~\cite{Bodlaender93}, being extremely popular in parameterized algorithmics.
Without the contributions of Hans Bodlaender, this would be 
much less so.

Intuitively, the fundamental observation behind treewidth is that 
many NP-hard graph problems turn easy when restricted to trees. 
Indeed, typically a simple bottom-up greedy algorithm 
from the leaves to the (arbitrarily chosen) root of the tree suffices to 
solve many fundamental graph problems (including \textsc{Vertex Cover}
and \textsc{Dominating Set}) efficiently on trees.
This naturally leads to the investigation on how ``tree-likeness'' of graphs 
helps to solve problems efficiently.
Fruitful results on this are provided by the concept of a tree decomposition and, 
correspondingly,
the treewidth of a graph:
if the treewidth is small, then otherwise NP-hard problems can 
be solved ``fast''.
Notably, the concepts of tree decomposition and treewidth are tightly 
connected to the existence of small graph separators 
(that is, vertex sets whose 
deletion partitions the graph into at least two connected components) 
that are arranged in a tree-like structure (see Section~\ref{sec:prelim} 
for formalities and an example). 
It is fair to say that tree decompositions currently are the 
most popular structural graph decompositions used in 
(parameterized) algorithms for (NP-hard) problems on (static) graphs.
More specifically, these algorithmic results typically are
``fixed-parameter tractability'' results with respect to the parameter,
that is, the studied problems then can be solved by an exponential-time
algorithm whose exponential part exclusively depends on 
the treewidth of the input graph.

Computing the treewidth of a graph is \NP-hard~\cite{ArnborgCP87}, even on graphs of maximum degree nine~\cite{BodlaenderT97},
but linear-time solvable if the treewidth is some fixed constant~\cite{Bodlaender96,BodlaenderK96} (more specifically, the running time is $c^{k^3} n$ \cite{Bodlaender96}) and 5-approximable in~$(c')^k n$~time~\cite{BodlaenderDDFLP16} on~$n$-vertex graphs of treewidth~$k$,
for some constants~$c, c'>1$. 
Polynomial-time algorithms are known for several restricted graph classes~\cite{BouchitteT01,BodlaenderM93,BodlaenderKK95,BodlaenderKKM98}.
From an algorithmic point of view,
a tree decomposition of a graph typically allows for a dynamic programming  approach.
The twist is that these dynamic programs for many \NP-hard problems run 
in polynomial time
when the width of the tree decomposition is constant~\cite{Bodlaender00}. 
Indeed,
many \NP-hard problems are known to be \fpt{} when parameterized by treewidth~\cite{DowneyF99,CyganFKLMPPS15},
underpinning the reputation of treewidth as one of the most fundamental algorithmically exploitable graph parameters, as confirmed
in experimental studies~\cite{BodlaenderK11,BodlaenderH98}
(practically useful implementations for computing tree decompositions are also available~\cite{DellKTW17,DellHJKKR16}).
In this work,
our goal is to discuss the role treewidth currently plays in the strongly
emerging field of \emph{temporal} graphs\footnote{Also known as time-varying
graphs, evolving graphs, link streams, or dynamic graphs where no changes on the
vertex set are allowed.}; these are graphs where the edge set may change over time.
Further, we reflect on possible definitions of a temporal version of treewidth: 
temporal treewidth.

Temporal graphs model networks where adjacencies of vertices change over discrete time steps.
In fact,
many natural time-dependent networks can be modeled by temporal graphs,
for instance interaction/contact networks,
connection/availability networks,
or bio-physical networks (see, e.g., \cite[Section~II]{HolmeS11}).
Applications range from epidemiology
over sociology
to transportation.
In the last decade,
problems on temporal graphs gained increased attention in theoretical computer science \cite{Akr+19,AxiotisF16,Ben+19,FluschnikMNRZ20,ErlebachKLSS19,ZschocheFMN18,FluschnikNRZ19,ErlebachS18,Erlebach0K15,AkridaMS19,BodlaenderZ19,CasteigtsPS19,FlocchiniMS13,VLM16,MMZ19,CasteigtsFQS12,FroeseJNR19,MolterNR19,HimmelBNN19}.

Formally,
a temporal graph~$\TG=(V,\TE,\tau)$ consists of a vertex set~$V$, 
a lifetime~$\tau$,
and a set~$\TE\subseteq \binom{V}{2}\times\set{\tau}$ of temporal edges 
(that is, an edge is additionally equipped with a \emph{time stamp}).
Alternatively,
a temporal graph on vertex set~$V$ and lifetime~$\tau$ can also be defined as (see~\cref{fig:tempgr} for exemplary illustrations)
\begin{compactenum}[(a)]
  \item a static graph~$G$ equipped with a function~$\lambda\colon E(G)\to 2^{\set{\tau}}$ (\Cref{fig:tempgr}(a)), 
  \item a tuple~$(V,E_1,\ldots,E_\tau)$ (\Cref{fig:tempgr}(b)), or 
  \item a sequence of~$\tau$ static graphs (called \emph{layers})~$G_1=(V,E_1),\ldots,G_\tau=(V,E_\tau)$ (\Cref{fig:tempgr}(c)). 
\end{compactenum}
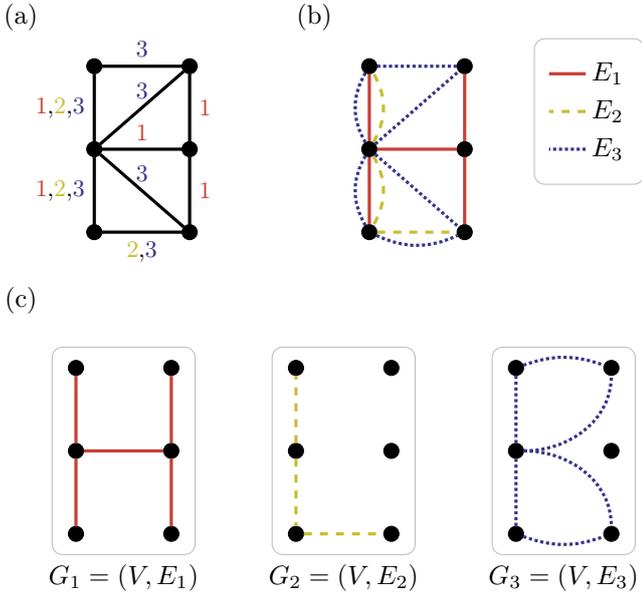
\begin{figure}[t]
 \centering
 
 \definecolor{layer1}{RGB}{200,63,53}
 \definecolor{layer2}{RGB}{200,190,53}
 \definecolor{layer3}{RGB}{55,52,139}
 \begin{tikzpicture}

  \usetikzlibrary{calc}
  \tikzstyle{xnode}=[fill,circle,draw,scale=0.6];
  \tikzstyle{xedge}=[very thick,-];
  \tikzstyle{xlabel}=[midway,scale=0.9];
  \tikzstyle{l1edge}=[xedge,layer1];
  \tikzstyle{l2edge}=[xedge,layer2,dashed];
  \tikzstyle{l3edge}=[xedge,layer3,densely dotted];

  \def\xr{0.965}
  \def\yr{1}
  \def\xsc{1.3}
  \def\ysc{1.1}

  \newcommand{\hlbnodes}[2]{
  \node (l2) at (#1,#2+1*\ysc*\yr)[xnode]{};
    \node (l1) at (#1,#2)[xnode]{};
  \node (l0) at (#1,#2-1*\ysc*\yr)[xnode]{};
  \node (r2) at (#1+1*\xsc*\xr,#2+1*\ysc*\yr)[xnode]{};
    \node (r1) at (#1+1*\xsc*\xr,#2)[xnode]{};
  \node (r0) at (#1+1*\xsc*\xr,#2-1*\ysc*\yr)[xnode]{};
  }

  \begin{scope}[xshift=\xr*0.25 cm]
  \node at (-1.0*\xr,1.75*\yr)[]{(a)};
    \hlbnodes{0}{0};
    \draw[xedge] (l1) to node[xlabel,left]{\textcolor{layer1}{1},\textcolor{layer2}{2},\textcolor{layer3}{3}}(l0);
    \draw[xedge] (l1) to node[xlabel,left]{\textcolor{layer1}{1},\textcolor{layer2}{2},\textcolor{layer3}{3}}(l2);

    \draw[xedge] (r1) to node[xlabel,right]{\textcolor{layer1}{1}}(r0);
    \draw[xedge] (r1) to node[xlabel,right]{\textcolor{layer1}{1}}(r2);

    \draw[xedge] (l1) to node[xlabel,above]{\textcolor{layer1}{1}}(r1);
    \draw[xedge] (l0) to node[xlabel,below]{\textcolor{layer2}{2},\textcolor{layer3}{3}}(r0);
    \draw[xedge] (l2) to node[xlabel,above]{\textcolor{layer3}{3}}(r2);

    \draw[xedge] (l1) to node[xlabel,above]{\textcolor{layer3}{3}}(r0);
    \draw[xedge] (l1) to node[xlabel,above]{\textcolor{layer3}{3}}(r2);
  \end{scope}

  \begin{scope}[xshift=\xr*4 cm]
  \node at (-0.75*\xr,1.75*\yr)[]{(b)};
    \hlbnodes{0}{0};
    \draw[l1edge] (l1) to (l0);
    \draw[l2edge] (l1) to [out=-60,in=60](l0);
    \draw[l3edge] (l1) to [out=-120,in=120](l0);
    \draw[l1edge] (l1) to (l2);
    \draw[l2edge] (l1) to [out=60,in=-60](l2);
    \draw[l3edge] (l1) to [out=120,in=-120](l2);

    \draw[l1edge] (r1) to (r0);
    \draw[l1edge] (r1) to (r2);

    \draw[l1edge] (l1) to (r1);
    \draw[l2edge] (l0) to (r0);
    \draw[l3edge] (l0) to [out=-25,in=205](r0);
    \draw[l3edge] (l2) to (r2);

    \draw[l3edge] (l1) to (r0);
    \draw[l3edge] (l1) to (r2);

    \begin{scope}[xshift=-\xr*0.75 cm,yshift=-\yr*0.5 cm]
        \node (e1) at (4*\xr,1.5*\yr)[]{$E_1$};
      \node (e2) at (4*\xr,1.0*\yr)[]{$E_2$};
      \node (e3) at (4*\xr,0.5*\yr)[]{$E_3$};
      \draw[l1edge] ($(e1)-(0.3*\xr,0)$) to ($(e1)-(0.8*\xr,0)$);
      \draw[l2edge] ($(e2)-(0.3*\xr,0)$) to ($(e2)-(0.8*\xr,0)$);
      \draw[l3edge] ($(e3)-(0.3*\xr,0)$) to ($(e3)-(0.8*\xr,0)$);
      \node at (3.75*\xr,1*\yr)[draw=lightgray,thin,rounded corners,minimum width=\xr*1.5 cm,minimum height=\xr*2 cm]{};
    \end{scope}
  \end{scope}

  \begin{scope}[yshift=-\yr*4.0cm]
  \node at (-0.75*\xr,2.0*\yr)[]{(c)};
    \hlbnodes{0}{0};
    \draw[l1edge] (l1) to (l0);
    \draw[l1edge] (l1) to (l2);
    \draw[l1edge] (r1) to (r0);
    \draw[l1edge] (r1) to (r2);
    \draw[l1edge] (l1) to (r1);
    \node at (0.5*\xsc*\xr,0*\yr)[draw=lightgray,thin,rounded corners,minimum width=\xr*\xsc*1.5 cm,minimum height=\yr*\ysc*2.5 cm,label=-90:{$G_1=(V,E_1)$}]{};

    \hlbnodes{3*\xr}{0};
    \draw[l2edge] (l1) to (l0);
    \draw[l2edge] (l1) to (l2);
    \draw[l2edge] (l0) to (r0);
      \node at (3*\xr+0.5*\xsc*\xr,0*\yr)[draw=lightgray,thin,rounded corners,minimum width=\xr*\xsc*1.5 cm,minimum height=\yr*\ysc*2.5 cm,label=-90:{$G_2=(V,E_2)$}]{};

    \hlbnodes{6*\xr}{0};
    \draw[l3edge] (l1) to (l0);
    \draw[l3edge] (l1) to (l2);
    \draw[l3edge] (l0) to [out=-20,in=-160](r0);
    \draw[l3edge] (l2) to [out=20,in=160](r2);
    \draw[l3edge] (l1) to [out=0,in=90](r0);
    \draw[l3edge] (l1) to [out=0,in=-90](r2);
    \node at (6*\xr+0.5*\xsc*\xr,0*\yr)[draw=lightgray,thin,rounded corners,minimum width=\xr*\xsc*1.5 cm,minimum height=\yr*\ysc*2.5 cm,label=-90:{$G_3=(V,E_3)$}]{};
  \end{scope}

  \end{tikzpicture}
  \caption{Three different illustrations (a)--(c) (according to alternative definitions (a)--(c)) of a temporal graph~$\TG$ with six vertices, fourteen temporal edges, and lifetime three.
  If the time stamps are dropped in~(a), the underlying graph is depicted.}
  \label{fig:tempgr}
\end{figure}
The \emph{underlying} (static) graph of a temporal graph~$\TG=(V,E_1,\ldots,E_\tau)$ is the graph~$\ug{G}(\TG)=(V,E_1\cup\dots\cup E_\tau)$.
If,
for instance,
contact networks are modeled,
then the underlying graph gives complete information on which contacts appeared;
however,
it gives no information on \emph{when} and \emph{how often} they appeared.

The addition of temporality to the graph model significantly increases the computational complexity of many basic graph problems.
For instance,
consider the following:
given a graph~$G$ with a designated vertex~$s$, 
decide whether there is a walk in~$G$ that starts at~$s$ and reaches all vertices.
(One may think of $s$ as the starting location of a traveling salesman who wants to visit every vertex at least once.)
This problem is well-known to be linear-time solvable on static graphs:
a solution exists if and only if $G$ is connected.
On temporal graphs,
the question is to decide whether all vertices can be reached from $s$ by a single so-called \emph{strict temporal walk},
that is, a walk whose sequence of temporal edges has strictly increasing time stamps. 
(Intuitively, strict temporal walks model the traversal of the graph at finite speed.)
This problem is known as \prob{Temporal Exploration}
and proven to be \NP-hard~\cite{AkridaMS19,BodlaenderZ19}.

When attempting to adapt the notion of treewidth to temporal graphs,
one might consider the \emph{underlying treewidth} $\utw(\TG)=\tw(\ug{G}(\TG))$.
Inherited from the loss of information in the underlying graph,
this notion captures no information about time and occurrences of temporal edges.
As we will see in~\cref{sec:intrac},
many temporal graph problems remain \NP-hard even when the underlying treewidth is constant,
indicating that the underlying treewidth is most probably missing useful ``time-structural'' information.
This lack of information seems to be even larger for the \emph{layer treewidth},
defined as~$\ltw(\TG):=\max_{i\in\set{\tau}}\tw(G_i)$ 
for a temporal graph~$\TG$ with layers~$G_1,\ldots,G_\tau$.
This is also expressed by the fact that~$\ltw(\TG)\leq \utw(\TG)$ for every temporal graph~$\TG$.

Nevertheless,
there are problems that are polynomial-time solvable on temporal graphs of constant underlying treewidth.
As we will see in~\cref{sec:tract},
several of these problems are actually \fpt{} when parameterized by the combination of the underlying treewidth and the lifetime.
Indeed,
in \cref{sec:mso},
we present a (temporal) adaption of the well-known technique of employing treewidth together with monadic second-order logic.
Again, we can obtain several
\fpty{} results when parameterizing by~$\utw+\tau$.

In search of a more useful definition of temporal treewidth,
we derive two requirements from observations in~\cref{sec:intrac,sec:tract}:
it should be at least as large as the underlying treewidth
and upper-bounded by some function in the combination of the underlying treewidth and the lifetime.
In~\cref{sec:reflections},
we will elaborate on this while reflecting on some possibly useful definitions for temporal treewidth.
All temporal graph problems encountered in this work are summarized in the appendix.

\section{Preliminaries}
\label{sec:prelim}

\casablanca{But what about us?} 
In this section, we provide some basic definitions and facts.
By~$\N$ and~$\N_0$ we denote the natural numbers excluding and including zero, 
respectively.
For any set $A$, we write $\binom{A}{k}$ for the set of all size-$k$ subsets from $A$.

\subsection{Static Graphs and Treewidth}\label{sec:TWDef}

Let~$G=(V,E)$ be a (static) graph with vertex set~$V$ and edge set~$E\subseteq \binom{V}{2}$.
Alternatively,  $V(G)$ and~$E(G)$ also denote the vertex set and edge set of~$G$,
respectively.
We write $G[W]$ for the subgraph induced by a set of vertices $W \subseteq V$ 
and use $G - W$ as a shorthand for $G[V\setminus W]$.

\paragraph{Tree Decompositions and Treewidth.}
In the following,
we define (rooted and nice) tree decompositions and treewidth of static graphs,
and we explain the connection to a cops-and-robber game.

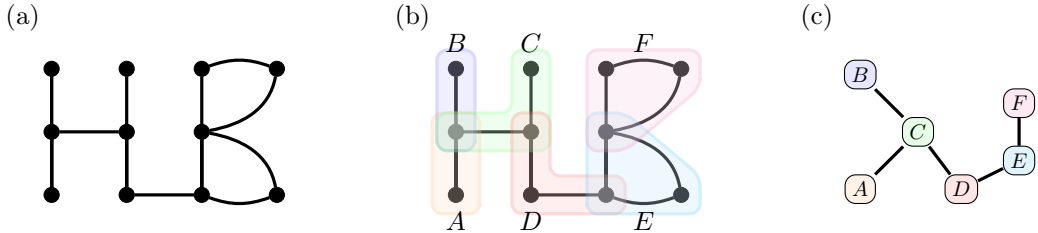
\begin{figure*}[t]
 \centering
 \begin{tikzpicture}
    \usetikzlibrary{calc}
    \def\xr{0.76}
    \def\yr{0.76}
    \def\xsc{1.3}
    \def\ysc{1.1}
    \def\xeps{0.25}
    \def\yeps{0.25}

    \tikzstyle{xnode}=[fill,circle,draw,scale=0.6];
    \tikzstyle{tdcnode}=[rectangle,rounded corners,draw,scale=0.8];
    \tikzstyle{xedge}=[very thick,-];
    \newcommand{\xbag}[2]{\draw[fill=#1!20!white,opacity=0.3,rounded corners,draw=#1!50!white, very thick] #2 to cycle;}

    \newcommand{\hlbnodesx}[2]{
      \node (l2) at (#1,#2+1*\ysc*\yr)[xnode]{};
      \node (l1) at (#1,#2)[xnode]{};
      \node (l0) at (#1,#2-1*\ysc*\yr)[xnode]{};
      \node (r2) at (#1+1*\xsc*\xr,#2+1*\ysc*\yr)[xnode]{};
      \node (r1) at (#1+1*\xsc*\xr,#2)[xnode]{};
      \node (r0) at (#1+1*\xsc*\xr,#2-1*\ysc*\yr)[xnode]{};
      \node (l2x) at (#1+2*\xsc*\xr,#2+1*\ysc*\yr)[xnode]{};
      \node (l1x) at (#1+2*\xsc*\xr,#2)[xnode]{};
      \node (l0x) at (#1+2*\xsc*\xr,#2-1*\ysc*\yr)[xnode]{};
      \node (r2x) at (#1+2*\xsc*\xr+1*\xsc*\xr,#2+1*\ysc*\yr)[xnode]{};
      \node (r0x) at (#1+2*\xsc*\xr+1*\xsc*\xr,#2-1*\ysc*\yr)[xnode]{};

      \draw[xedge] (l2) -- (l0) -- (l1);
      \draw[xedge] (l2x) -- (l0x) -- (l1x);
      \draw[xedge] (l1) -- (r1);
      \draw[xedge] (r2) -- (r0) -- (r1);
      \draw[xedge] (r0) -- (l0x);
      \draw[xedge] (l2) -- (l0) -- (l1);
      \draw[xedge] (l0x) to [out=-20,in=-160](r0x);
      \draw[xedge] (l2x) to [out=20,in=160](r2x);
      \draw[xedge] (l1x) to [out=-10,in=100](r0x);
      \draw[xedge] (l1x) to [out=10,in=-100](r2x);
      }

    \begin{scope}[xshift=\xr*0cm]
      \node at (-0.5*\xr,2*\yr)[]{(a)};
      \hlbnodesx{0}{0};
    \end{scope}

    \def\xsh{6}
    \ifdoublecolumn\def\xsh{7}\fi
    \begin{scope}[xshift=\xr*\xsh cm]
      \node at (-0.75*\xr,2*\yr)[]{(b)};
      \hlbnodesx{0}{0};
      \xbag{orange}{($(l0)+(1.25*\xeps,-\yeps)$) 
      to ($(l1)+(1.25*\xeps,\yeps)$)
      to ($(l1)+(-1.25*\xeps,\yeps)$)
      to ($(l0)+(-1.25*\xeps,-\yeps)$) 
      }
      \node at (l0.south)[anchor=north]{$A$};
      \xbag{blue}{
      ($(l1)+(\xeps,-\yeps)$) 
      to ($(l2)+(\xeps,\yeps)$)
      to ($(l2)+(-\xeps,\yeps)$)
      to ($(l1)+(-\xeps,-\yeps)$) 
      }
      \node at (l2.north)[anchor=south]{$B$};
      \xbag{green}{
      ($(l1)+(-\xeps,-\yeps)$) 
      to ($(r1)+(\xeps,-\yeps)$)
      to ($(r2)+(\xeps,\yeps)$)
      to ($(r2)+(-\xeps,\yeps)$) 
      to ($(r1)+(-\xeps,\yeps)$) 
      to ($(l1)+(-\xeps,\yeps)$) 
      }
      \node at (r2.north)[anchor=south]{$C$};
      \xbag{red}{
      ($(r1)+(-\xeps,-\yeps)$) 
      to ($(r0)+(-\xeps,-\yeps)$)
      to ($(l0x)+(\xeps,-\yeps)$)
      to ($(l0x)+(\xeps,\yeps)$) 
      to ($(r0)+(\xeps,\yeps)$) 
      to ($(r1)+(\xeps,\yeps)$) 
      to ($(r1)+(-\xeps,\yeps)$) 
      }
      \node at (r0.south)[anchor=north]{$D$};
      \xbag{cyan}{
      ($(l0x)+(-\xeps,-\yeps)$) 
      to ($(r0x)+(\xeps,-\yeps)$)
      to ($(r0x)+(\xeps,\yeps)$)
      to ($(l1x)+(\xeps,\yeps)$) 
      to ($(l1x)+(-\xeps,\yeps)$) 
      }
      \draw[draw=none] (l0x) to node[below,yshift=-\yr*0.125cm]{$E$}(r0x);
      \xbag{magenta}{
      ($(r2x)+(\xeps,-\yeps)$)
      to ($(l1x)+(\xeps,-\yeps)$) 
      to ($(l1x)+(-\xeps,-\yeps)$) 
      to ($(l2x)+(-\xeps,\yeps)$) 
      to ($(l2x)+(\xeps,\yeps)$) 
      to ($(r2x)+(\xeps,\yeps)$)
      }
      \draw[draw=none] (l2x) to node[above,yshift=\yr*0.125cm]{$F$}(r2x);
    \end{scope}

    \def\xsh{13}
    \ifdoublecolumn\def\xsh{15}\fi
    \begin{scope}[xshift=\xr*\xsh cm]
      \node at (-1.75*\xr,2*\yr)[]{(c)};
      \node (c) at (0,0)[tdcnode,fill=green!10!white]{$C$};
      \node (a) at (-1*\xr,-1*\yr)[tdcnode,fill=orange!10!white]{$A$};
      \node (b) at (-1*\xr,1*\yr)[tdcnode,fill=blue!10!white]{$B$};
      \node (d) at (0.75*\xr,-1*\yr)[tdcnode,fill=red!10!white]{$D$};
      \node (e) at (1.75*\xr,-0.5*\yr)[tdcnode,fill=cyan!10!white]{$E$};
      \node (f) at (1.75*\xr,0.5*\yr)[tdcnode,fill=magenta!10!white]{$F$};
      \draw[xedge] (a) -- (c) -- (b);
      \draw[xedge] (c) -- (d) -- (e) -- (f);
    \end{scope}
    \end{tikzpicture}
    \caption{(a) A graph~$G$ of treewidth two with (c) a tree decomposition of~$G$ with (b) bags~$A$--$F$.}
     \label{fig:tdc}
\end{figure*}

\begin{definition}[Tree Decomposition, Treewidth]
 \label{def:treedecwidth}
 Let~$G=(V,E)$ be an undirected graph.
 Then a tuple~$\bbT=(T,\{B_u\mid u\in V(T)\})$ consisting of a tree~$T$ and a set of so-called bags~$B_u\subseteq V$ is a \emph{tree decomposition (\tdc{})} of~$G$ if
 \begin{enumerate}[(i)]
  \item $\bigcup_{u\in V(T)} B_u  = V$,
  \item for every~$e\in E$ there is a node~$u\in V(T)$ such that~$e\subseteq B_u$, and
  \item for every~$v\in V$, the graph~$T[\{u\in V(T)\mid v\in B_u\}]$ is a tree.
 \end{enumerate}
 The \emph{width} of~$\bbT$ is~$\width(\bbT):=\max_{u\in V(T)}|B_u|-1$.
 The treewidth of~$G$ is the minimum width over all tree decompositions of~$G$,
 that is,  
 \[\tw(G)=\min\limits_{\bbT\text{ is \tdc{} of }G} \width(\bbT). \]
\end{definition}
It follows from the definition, that for any edge $\{u, u'\}$ of $T$, the intersection of the corresponding bags $B_u \cap B_{u'}$ is a separator of $G$ of size at most $\width(\bbT)$
(as long as $\bbT$ does not contain redundant bags).
Refer to~\cref{fig:tdc} for an illustrative example.

A tree decomposition~$\bbT=(T,\{B_u\mid u\in V(T)\})$ is \emph{rooted}
if there is a designated node~$r\in V(T)$ being the root of~$T$ 
(this allows to talk about children, parents, ancestors, descendants, etc.\ of the nodes of~$T$).
A rooted tree decomposition~$\bbT=(T,\{B_u\mid u\in V(T)\})$ is \emph{nice} if each node~$u\in V(T)$ is either
\begin{inparaenum}[(i)]
 \item a leaf node ($u$ has no children),
 \item an introduce node ($u$ has one child~$v$ with~$B_v\subset B_u$ and~$|B_u\setminus B_v|=1$),
 \item a forget node ($u$ has one child~$v$ with~$B_v\supset B_u$ and~$|B_v\setminus B_u|=1$), or
 \item a join node ($u$ has two children~$v,w$ with~$B_v=B_w=B_u$).
\end{inparaenum}
Given a tree decomposition, one can compute a corresponding nice tree decomposition in linear time~\cite{Kloks94}.

Alternatively,
treewidth can be defined through a cops-and-robber game~\cite{SeymourT93}
as follows.
Let~$G=(V,E)$ be an undirected graph, 
and~$k\in\N$.
\begin{itemize}
 \item At the start,
the~$k$ cops choose a set~$C_0\in\binom{V}{k}$ of vertices,
and then the robber chooses a vertex~$r_0\in V\setminus C_0$.
  \item In round~$i\in\N$,
first the cops choose~$C_i\in \binom{V}{k}$,
and then the robber chooses~$r_i\in V\setminus C_i$ such that 
$r_i$ and~$r_{i-1}$ are connected in~$G-(C_i\cap C_{i-1})$.
\end{itemize}
The cops win if, after finitely many rounds,
the robber is caught, 
that is,
the robber has no vertex left to choose.
The connection to treewidth is the following 
(which also implies an alternative definition for treewidth).

\begin{lemma}[\cite{SeymourT93}]
 \label{lem:twcar}
 Graph~$G$ has treewidth at most~$k$ if and only if at most~$k+1$ cops win the cops-and-robber game.
\end{lemma}

The \emph{pathwidth} of a graph~$G$ is the minimum width over all tree decomposition~$\bbT=(T,\{B_u\mid u\in V(T)\})$ with~$T$ being a path.
Note that for every graph its treewidth is at most its pathwidth.
For the graph in~\cref{fig:tdc},
the treewidth and the pathwidth are equal
(take the union of the bags~$A$ and~$B$).

\subsection{Temporal Graphs}
Let $\TG=(V,\TE,\tau)$ be a temporal graph (see \cref{sec:intro}).
We also denote by~$V(\TG)$ the vertex set of~$\TG$.
For any vertex subset~$W\subseteq V$,
the temporal graph~$\TG[W]$ induced by~$W$ is defined as~$(W,\{(e,t)\in\TE\mid e\subseteq W\},\tau)$.
Further, we define $\TG - W :=\TG[V\setminus W]$.
For a subset of temporal edges~$\TE'\subseteq \TE$,
the temporal graph~$\TG-\TE'$ is defined as~$(V,\TE\setminus\TE',\tau)$.

A \emph{temporal walk} is defined as a sequence of temporal edges~$(\{v_1,v_{2}\},t_1),\allowbreak(\{v_2,v_{3}\},t_2),\allowbreak\dots,\allowbreak(\{v_p,v_{p+1}\},t_p)$,
each contained in~$\TE$ and~$t_1\leq t_2\leq \dots\leq t_p$ (also called a $v_1$\nobreakdash-$v_{p+1}$~temporal walk when the terminals are specified).
A temporal walk is called \emph{strict} if $t_1< t_2<\dots< t_p$.
A \emph{(strict) temporal path} is a (strict) temporal walk where $v_i\neq v_j$ for all~$i\neq j$.
A (strict)~$(\alpha,\beta)$-temporal path is a (strict) temporal path where additionally~$\alpha\leq t_{i+1}-t_i\leq \beta$ holds.

When analyzing problems on temporal graphs, the following concept often comes in handy.
\begin{definition}[(Strict) Static Expansion]
 \label{def:strstaticexp}
 The \emph{static expansion} of a temporal graph~$\TG=(V,\TE,\tau)$ is a directed graph~$H := (V',A)$, 
with vertices $V' = \{ u_{t,j} \mid v_j\in V ,\, t\in\set{\tau} \}$ and
arcs $A = A' \cup A_{\rm col}$, where the first set
    $A':=\{ ( u_{t,i},u_{t,i'}), (u_{t,i'}, u_{t,i}) \mid (\{ v_i,v_{i'}\},t) \in \TE\}$
contains the arcs within the layers, and the second set
    $A_{\rm col}:=\{ (u_{t,j},u_{t+1,j}) \mid v_j\in V ,\, t\in\set{\tau-1}\}$
contains the arcs connecting different layers.
We refer to~$A_{\rm col}$ as \emph{column-edges} of~$H$.
 
 A static expansion is called \emph{strict} if its vertex set~$V'$ additionally contains the vertex set~$\{ u_{\tau+1,j} \mid v_j\in V\}$
 and its arc set~$A'$ is replaced by the set~$A'':=\{ ( u_{t,i},u_{t+1,i'}), (u_{t,i'}, u_{t+1,i}) \mid (\{ v_i,v_{i'}\},t) \in \TE\}$.
\end{definition}
Note that (strict) temporal walks correspond exactly to walks within the (strict) static expansion.
Moreover, note that \emph{strict} static expansions are always directed \emph{acyclic} graphs.

\subsection{Parameterized Complexity}

We use standard notation and terminology from parameterized
complexity theory~\cite{DowneyF99,Downey2013OtherWidthMetrics,FG06,Nie06,CyganFKLMPPS15}.
A parameterized problem with parameter $k$ is a language~$L\subseteq \{(x,k)\in \Sigma^*\times\N\}$ for some finite alphabet~$\Sigma$.
A parameterized problem~$L$ is called \emph{\fpt{}} if every instance~$(x,k)$ can be decided for~$L$ in~$f(k)\cdot |x|^{\bigO(1)}$ time,
where~$f$ is some computable function only depending on~$k$.
The tool for proving that a parameterized problem is presumably not \fpt{} is to show that it is hard for the parameterized complexity class \W{1}. 
A parameterized problem~$L$ is contained in the complexity class~$\XP$ if every instance~$(x,k)$ can be decided for~$L$ in~$|x|^{g(k)}$ time,
where $g$ is some computable function only depending on~$k$.
A parameterized problem~$L$ is \emph{para-\NP-hard} if the problem is \NP-hard for some constant value of the parameter.

\section{Intractability for Constant Underlying Treewidth}
\label{sec:intrac}
\casablanca{I’m saying this because it’s true.}
For static graphs,
many \NP-hard problems become polynomial-time solvable when the input graph has constant treewidth.
More specifically, 
many such problems are \fpt{} when parameterized by treewidth;
comparatively few problems are \W{1}-hard yet contained in~\XP{} when parameterized by treewidth~\cite{BBNU12,DomLSV08,FluschnikKNS19,DvorakK18}
or remain \NP-hard when restricted to graphs of constant treewidth~\cite{Marx05,Marx09,NishizekiVZ01,Gassner10}.
For temporal graphs,
parametrization by the underlying treewidth leads to quite different observations:
so far, few temporal problems are known to be contained in~\XP{} or even \fpt{},
while several problems remain \NP-hard even if the underlying treewidth is constant (see~\cref{tab:results} for an overview of the results and \cref{sec:probzoo} for problem definitions).
\begin{table*}[t]
 \caption{Overview on treewidth-related results for \NP-hard temporal graph problems.
 \ifdoublecolumn{}\else{}%
 The problems are \prob{(Return-To-Base) Temporal Graph Exploration ((RTB-)TGE)},
 \prob{$(\alpha,\beta)$-Temporal Reachability Time-Edge Deletion ($(\alpha,\beta)$-TRTED)},
 \prob{Reachability Temporal Ordering (RTO)},
 \prob{Min Reachability Temporal Merging (MRTM)},
 \prob{Temporal Matching (TM)},
 \prob{Temporal Separation (TS)}, 
 and
 \prob{Minimum Single-Source Temporal Connectivity ($r$-MTC)}.
 \fi{}%
 See~\cref{sec:probzoo} for respective problem definitions. 
 ``FPT'', ``p-\NP-h'', and ``?'' abbreviate ``\fpt{}'', ``para-\NP-hard'', and ``open'', respectively.\newline
 $^\dagger$\,Results not (explicitly) contained in the given literature reference (last column) yet being simple observations/corollaries.
 $^\star$\,Open whether contained in \FPT.
}
 \label{tab:results}
 \centering
 \setlength{\tabcolsep}{7pt}
 
 \ifdoublecolumn{}
 \setlength{\tabcolsep}{5.5pt}
  \begin{tabular}{@{}lll|l|ll|l@{}}\toprule
  \diagbox[trim=lr]{Problem}{Parameter} & \ltw{} & \utw{} & $\tau$ & $|V|$ & $\utw+\tau$ & Ref.\\\midrule\midrule
  \prob{Temporal Graph Exploration}                    & \multicolumn{2}{c|}{\NP-h for $\utw=2$} & \FPT$^\dagger$ & \FPT$^\dagger$ & \FPT$^\dagger$ & \cite{BodlaenderZ19} \\
  \prob{Return-To-Base Temporal Graph Exploration}                & \multicolumn{2}{c|}{\NP-h for $\utw=1$} & \FPT$^\dagger$ & \FPT$^\dagger$ & \FPT$^\dagger$ & \cite{AkridaMS19} \\
  \prob{$(\alpha,\beta)$-Temporal Reachability Time-Edge Deletion} & \multicolumn{2}{c|}{\NP-h for $\utw=1$} & p-\NP-h & ? & ? & \cite{EnrightMMZ18} \\
  \prob{$\binom{\text{Min-Max}}{\text{Max-Min}}$ Reachability Temporal Ordering}            & \multicolumn{2}{c|}{\NP-h for $\utw=1$} & \FPT{} & ? & \FPT{} & \cite{EnrightMS18}\\
  \prob{Min Reachability Temporal Merging}                   & \multicolumn{2}{c|}{\NP-h for $\utw=1$} & \FPT$^\dagger$ & ? & \FPT$^\dagger$ & \cite{DeligkasmP19} \\
  \prob{Temporal Matching}                     & \multicolumn{2}{c|}{\NP-h for $\utw=1$}  & p-\NP-h & ? & \FPT$^\dagger$ & \cite{MertziosMNZZ20} \\
  \midrule
  \prob{Temporal Separation}                     & p-\NP-h & \XP$^\star$  & p-\NP-h & \FPT{} & \FPT{} & \cite{ZschocheFMN18,FluschnikMNRZ20} \\
  \prob{Minimum Single-Source Temporal Connectivity}                     & p-\NP-h$^\dagger$ & \XP$^\star$ & ? & ? & \FPT & \cite{AxiotisF16} \\
  \bottomrule
 \end{tabular}
 \else{}
 \begin{tabular}{@{}lll|l|ll|l@{}}\toprule
  \diagbox[trim=lr]{Problem}{Parameter} & \ltw{} & \utw{} & $\tau$ & $|V|$ & $\utw+\tau$ & Ref.\\\midrule\midrule
  \prob{TGE}                    & \multicolumn{2}{c|}{p-\NP-h ($\utw=2$)} & \FPT$^\dagger$ & \FPT$^\dagger$ & \FPT$^\dagger$ & \cite{BodlaenderZ19} \\
  \prob{RTB-TGE}                & \multicolumn{2}{c|}{p-\NP-h ($\utw=1$)} & \FPT$^\dagger$ & \FPT$^\dagger$ & \FPT$^\dagger$ & \cite{AkridaMS19} \\
  \prob{$(\alpha,\beta)$-TRTED} & \multicolumn{2}{c|}{p-\NP-h ($\utw=1$)} & p-\NP-h & ? & ? & \cite{EnrightMMZ18} \\
  \prob{$\binom{\text{Min-Max}}{\text{Max-Min}}$ RTO}            & \multicolumn{2}{c|}{p-\NP-h ($\utw=1$)} & \FPT{} & ? & \FPT{} & \cite{EnrightMS18}\\
  \prob{MRTM}                   & \multicolumn{2}{c|}{p-\NP-n ($\utw=1$)} & \FPT$^\dagger$ & ? & \FPT$^\dagger$ & \cite{DeligkasmP19} \\
  \prob{TM}                     & \multicolumn{2}{c|}{p-\NP-h ($\utw=1$)}  & p-\NP-h & ? & \FPT$^\dagger$ & \cite{MertziosMNZZ20} \\
  \midrule
  \prob{TS}                     & p-\NP-h & \XP$^\star$  & p-\NP-h & \FPT{} & \FPT{} & \cite{ZschocheFMN18,FluschnikMNRZ20} \\
  \prob{$r$-MTC}                     & p-\NP-h$^\dagger$ & \XP$^\star$ & ? & ? & \FPT & \cite{AxiotisF16} \\
  \bottomrule
 \end{tabular}
 \fi{}
\end{table*}

Next, we try to understand a little better why constraining the parameter 
underlying treewidth seems to offer so little algorithmic benefit.
The reductions proving \NP-hardness on constant-treewidth underlying graphs have the following features in common:
\begin{itemize}
 \item the constructed underlying graph is tree-like, 
 \item vertices and time stamps capture structures of the input instance, and hence
 \item the numbers of vertices and layers are unbounded.
\end{itemize}
In the remainder of this section we present some concrete example 
reductions in moderate detail.
Our selected \NP-hardness reductions cover problems
from the fields of 
temporal exploration (\cref{ssec:hardnessTE}),
temporal reachability (\cref{ssec:hardnessTR}),
and temporal matching (\cref{ssec:hardnessTM}).

\subsection{Temporal Exploration}
\label{ssec:hardnessTE}

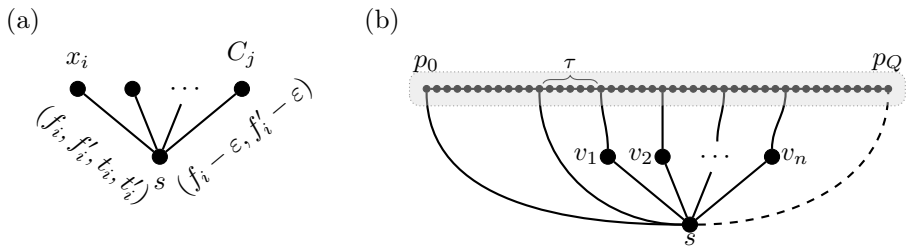
\begin{figure*}[t!]
 \centering
 \begin{tikzpicture}
 
  \tikzstyle{xnode}=[circle, fill=black, draw, scale=0.6]
  \tikzstyle{xedge}=[thick,-]

  \def\xr{0.9}
  \def\yr{0.9}
  
  \def\lvdist{1}
  \def\lhdist{0.4}
  \newcommand{\astar}[5]{
    \node (#3center) at (#1,#2)[xnode]{};
    \node (#3l1) at (#1-3*\lhdist*\xr,#2+#4*\lvdist*\yr)[xnode]{};
    \node (#3l2) at (#1-\lhdist*\xr,#2+#4*\lvdist*\yr)[xnode]{};
    \node (#3li) at (#1+\lhdist*\xr,#2+#4*\lvdist*\yr)[]{#5};
    \node (#3ln) at (#1+3*\lhdist*\xr,#2+#4*\lvdist*\yr)[xnode]{};
    \foreach \x in {1,2,i,n}{\draw[xedge] (#3center) to (#3l\x);}
  }
  \begin{scope}[yshift=\yr*1cm] 
    \node at (-2*\xr,2*\yr)[]{(a)};
    \astar{0}{0}{}{1}{$\cdots$};
    \node[below =of center,yshift=\yr*3em] (a) []{$s$};
    \node[above =of l1,yshift=-\yr*3em] (a) []{$x_i$};
    \node[above =of ln,yshift=-\yr*3em] (a) []{$C_j$};
    \node[above =of l1,yshift=-\yr*7em] (a) [rotate=-40]{$(f_i,f_i',t_i,t_i')$};
    \node[above =of ln,xshift=\xr*0.75em,yshift=-\yr*6.25em] (a) [rotate=40]{$(f_i-\eps,f_i'-\eps)$};
  \end{scope}

  \begin{scope}[xshift=\xr*7.75cm]
    \node at (-4.5*\xr,3*\yr)[]{(b)};
    \astar{0}{0}{}{1}{$\cdots$};

    \foreach \x in {1,2,...,46}{
      \node (p\x) at (-4*\xr+0.15*\xr*\x,2*\yr)[xnode,scale=0.45]{};
    }
    \draw[xedge] (p1) to (p46);
    
    \draw[xedge] (center) to [out=180,in=-90](p1);
    \draw[xedge] (center) to [out=180,in=-90](p12);
    \draw[xedge] (l1) to [out=90,in=-90](p18);
    \draw[xedge] (l2) to [out=90,in=-90](p24);
    \draw[xedge] (li) to [out=90,in=-90](p30);
    \draw[xedge] (ln) to [out=90,in=-90](p36);
    \draw[xedge,dashed] (center) to [out=0,in=-90](p46);
    
    \node[above =of p1,yshift=-\yr*3em] (a){$p_0$};
    \node[above =of p46,yshift=-\yr*3em] (a){$p_Q$};
    \node[left=of l1,xshift=\xr*3.5em] (a){$v_1$};
    \node[left=of l2,xshift=\xr*3.5em] (a){$v_2$};
    \node[right=of ln,xshift=-\xr*3.5em] (a){$v_n$};
    \node[below=of center,yshift=\yr*3.5em] (a){$s$};
    
    \draw[decorate,decoration={brace,amplitude=3pt,raise=2pt},yshift=\yr*10 em] (p12) -- (p18) node [above,midway,yshift=\yr*0.5 em] {\footnotesize $\tau$};%
    \draw[rounded corners, densely dotted, fill=gray!30!white,opacity=0.4] ($(p1)-(0.25*\xr,0.25*\yr)$) rectangle  ($(p46)+(0.25*\xr,0.25*\yr)$);
  \end{scope}
  \end{tikzpicture}
  \caption{Illustration of reductions behind~\cref{thm:rtbtge} (i) and (ii). %
 (a) A star with leaves corresponding to clauses and variables~\cite{AkridaMS19}. 
 (b) A star with center~$s$ and leaves~$v_1,\dots,v_n$; a path (highlighted in gray) on~$Q+1$ vertices is attached to the star~\cite{BodlaenderZ19}.
 }
 \label{fig:hardnessA}
\end{figure*}

In the problem called \prob{Return-To-Base Temporal Graph Exploration (RTB-TGE)},
one is given a temporal graph~$\TG$ and a designated vertex~$s$,
and the task is to decide whether there is a strict temporal walk starting and ending at~$s$ that visits all vertices in~$V(\TG)$.
The \NP-hardness of \prob{RTB-TGE} follows by a simple reduction from \prob{Hamiltonian Cycle (HC)}:
given a directed graph~$G$,
decide whether there is a (simple) cycle in~$G$ that contains all vertices from~$G$.
However,
from a parameterized view regarding the (underlying) treewidth,
\prob{RTB-TGE} is much harder than \prob{HC}:
while \prob{HC} parameterized by treewidth is \fpt{},
for \prob{RTB-TGE} we have the following.
\begin{theorem}[\cite{AkridaMS19,BodlaenderZ19}]
 \label{thm:rtbtge}
 \prob{Return-To-Base Temporal Graph Exploration} is \NP-hard even if 
 \begin{compactenum}[(i)]
  \item the underlying graph is a star or\label{rtbtge:akrida}
  \item each layer is a tree and the underlying graph has pathwidth at most two.\label{rtbtge:bodlaender}
 \end{compactenum}
\end{theorem}
\citet{AkridaMS19} proved \xref{thm:rtbtge}{rtbtge:akrida} via a reduction from \prob{3-SAT(3)}, 
a special case of \prob{3-SAT} where each variable appears in at most three clauses.
See \cref{fig:hardnessA}(a) for an illustration.
In the reduction from \prob{3-SAT(3)},
a star is constructed where 
for each variable and clause in the input \prob{3-SAT} formula
there is a leaf in the star.
Moreover, each variable~$x_i$ has two unique entry time steps~$f_i,t_i$ and two unique exit time steps~$f_i',t_i'$,
corresponding to setting $x_i$ to false (entering at~$f_i$ and leaving at~$f_i'$) or true (entering at~$t_i$ and leaving at~$t_i'$)
with $f_i < f_i' < t_i < t_i'$.
Clearly, it is never beneficial for the explorer to linger in a leaf longer than necessary.
Now assume clause~$C_j$ to contain variable~$x_i$ unnegated.
Then we add to~$C_j$ an entry time step~$f_i-\eps$ and an exit time step~$f_i'-\eps$.
Since $f_i - \eps < f_i < f_i' - \eps < f_i'$, clause~$C_j$ can be visited at time $f_i-\eps$ if and only if $x_i$ is set to true.

By adding analogous entry and exit time steps to~$C_j$ for all its contained variables (negated or unnegated),
it follows that $C_j$ can be visited if and only if at least one of its literals is set to true.

\citet{BodlaenderZ19} proved \xref{thm:rtbtge}{rtbtge:bodlaender} via a reduction from \prob{RTB-TGE} with the underlying graph being a star to \prob{(RTB-)TGE} by adding a long path to each layer, 
which is connected to some of the star's leaves in such a way
that each layer is a tree
and 
the underlying graph has pathwidth at most~two. 
See \cref{fig:hardnessA}(b) for an illustration.
In the reduction from \prob{RTB-TGE},
let~$\TG$ be the temporal graph with lifetime~$\tau$ and the underlying graph being a star on vertices~$s$ and~$v_1,\ldots,v_n$.
Then a temporal graph~$\TG'$ with lifetime~$\tau'=Q+\tau+1$ is constructed from~$\TG$
by adding a path on vertices~$p_0,\ldots,p_Q$ (highlighted in gray and present in all layers),
where $Q=\tau\cdot (n+4)$,
and appending $Q+1$ layers, in which each vertex~$v_i$ is adjacent only to $s$.
Furthermore, in each of the first~$\tau$ layers,
each vertex~$v_i$ is connected to some vertex on the path~$P$ if and only if $v_i$ is the lowest numbered vertex in a connected component.
This guarantees that every layer is connected.
Equivalence holds since in the first~$\tau$ time steps,
$\TG$ must be explored, 
and in the remaining~$Q+1$ time steps the path~$P$ must be explored ``in one run'' 
(starting from~$s$, going to~$p_0$ and ending at~$p_Q$).
In the return-to-base variant, 
the exploration ends with stepping from~$p_Q$ to~$s$.

\subsection{Temporal Reachability}
\label{ssec:hardnessTR}

While the problem of temporal exploration asks whether a single agent can traverse the entire graph,
temporal reachability problems are concerned with the set of vertices reachable by an infinite number of agents, 
all starting simultaneous at a single vertex.
Clearly, for any given start vertex this set can be determined by a simple search tree on the static expansion.
As this setting can be understood as a model for information flow or disease spreading,
a natural question is how far the set of reachable vertices can be decreased using a limited number of graph modifications;
this can be understood as a measure of temporal graph connectivity.
In the following, 
we address this question for three different types of modification operations: 
deletion of time-edges, reordering of layers, and merging of layers.

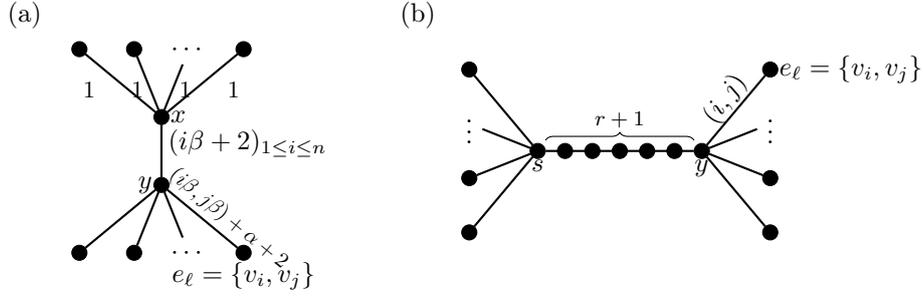
\begin{figure*}[t!]
 \centering
 \begin{tikzpicture}
 
  \tikzstyle{xnode}=[circle, fill=black, draw, scale=0.6]
  \tikzstyle{xedge}=[thick,-]

  \def\xr{0.9}
  \def\yr{0.9}
  
  \def\lvdist{1}
  \def\lhdist{0.4}
  \newcommand{\astar}[5]{
    \node (#3center) at (#1,#2)[xnode]{};
    \node (#3l1) at (#1-3*\lhdist*\xr,#2+#4*\lvdist*\yr)[xnode]{};
    \node (#3l2) at (#1-\lhdist*\xr,#2+#4*\lvdist*\yr)[xnode]{};
    \node (#3li) at (#1+\lhdist*\xr,#2+#4*\lvdist*\yr)[]{#5};
    \node (#3ln) at (#1+3*\lhdist*\xr,#2+#4*\lvdist*\yr)[xnode]{};
    \foreach \x in {1,2,i,n}{\draw[xedge] (#3center) to (#3l\x);}
  }

  \begin{scope}[xshift=\xr*0cm,yshift=-\yr*1.75cm]
    \node at (-2*\xr,1.5*\yr)[]{(a)};
    \astar{0}{0}{}{1}{$\cdots$};	
    \astar{0}{-1*\yr}{2}{-1}{$\cdots$};
    \draw[xedge] (center) to (2center);
     \node[right=of center,xshift=-\xr*3.5em] (a){$x$};
     \node[above=of center,yshift=-\yr*3em,xshift=-\xr*3em] (a)[scale=0.9]{$1$};
     \node[above=of center,yshift=-\yr*3em,xshift=-\xr*1em] (a)[scale=0.9]{$1$};
     \node[above=of center,yshift=-\yr*3em,xshift=\xr*1em] (a)[scale=0.9]{$1$};
     \node[above=of center,yshift=-\yr*3em,xshift=\xr*3em] (a)[scale=0.9]{$1$};
     \node[below right=of center,xshift=-\xr*3.5em,yshift=\yr*3.25em] (a){$(i\beta +2)_{1\leq i\leq n}$};
     \node[left=of 2center,xshift=\xr*3.5em] (a){$y$};
     \node[below=of 2ln,yshift=\yr*3.5em,xshift=-\xr*0.0em] (a){$e_\ell=\{v_i,v_j\}$};
     \node[above=of 2ln,xshift=-\xr*1.1em,yshift=-\yr*2.65em,scale=0.8] (a)[rotate=-39.5]{$(i\beta,j\beta)+\alpha+2$};
  \end{scope}

  \begin{scope}[xshift=\xr*5.5cm,yshift=-\yr*2.25cm,rotate=90]
    \node at (2*\xr,1.75*\yr)[]{(b)};
    \astar{0}{0}{}{1}{$\vdots$};	
    \astar{0}{-2.4*\yr}{2}{-1}{$\vdots$};
    \foreach \x in {1,2,...,5}{\node at (0,-0.4*\x*\yr)[xnode]{};}
    \draw[xedge] (center) to (2center);
    \node[below=of center,yshift=\yr*3.5em] (a){$s$};
    \node[below=of 2center,yshift=\yr*3.5em] (a){$y$};
    \node[right=of 2ln,xshift=-\xr*3.5em] (a){$e_\ell=\{v_i,v_j\}$};
    \node[left=of 2ln,xshift=\xr*2.5em] (a)[rotate=55]{$(i,j)$};
    \draw[decorate,decoration={brace,amplitude=3pt,raise=4pt},yshift=\yr*10 em] (center) -- (2center) node [above,midway,yshift=\yr*0.6 em] {\footnotesize $r+1$};
  \end{scope}
  \end{tikzpicture}
 \caption{Illustrations of the reductions behind \cref{thm:abTRTED,thm:mmrto}. %
 (a) Two stars with centers~$x$ and~$y$ and edge~$\{x,y\}$; The leaves of the star centered at~$y$ one-to-one correspond to the edges of the input graph (same for (b))~\cite{EnrightMMZ18}.
 (b) Two stars with centers~$s$ and~$y$, where the centers are connected via a path of~$r+1$ vertices~\cite{EnrightMS18}.
 }
 \label{fig:hardnessB}
\end{figure*}

\paragraph{Deletion of Time-Edges.}
\citet{EnrightMMZ18} studied the \prob{$(\alpha,\beta)$-Temporal Reachability Time-Edge Deletion ($(\alpha,\beta)$-TRTED)} problem:
given a temporal graph~$\TG=(V,\TE,\tau)$ and two integers~$k,h\geq0$,
decide whether there is a subset~$\TE'\subseteq \TE$ of temporal edges with~$|\TE'|\leq k$ such that in~$\TG-\TE'$,
the size of the set of vertices reachable from every vertex~$s\in V$ via strict~$(\alpha,\beta)$-temporal paths is at most~$h$.
\citeauthor{EnrightMMZ18} proved the following hardness result.
\begin{theorem}[\cite{EnrightMMZ18}]
 \label{thm:abTRTED}
 \prob{$(\alpha,\beta)$-Temporal Reachability Time-Edge Deletion} is \NP-hard
  even if the underlying graph consists of two stars with adjacent centers. 
\end{theorem}
The proof of \cref{thm:abTRTED} employs a reduction from \prob{Clique}:
given an undirected graph~$G$ and an integer~$r$,
decide whether~$G$ contains a clique (a graph where each pair of vertices is adjacent) with~$r$ vertices.
In the corresponding construction,
adjacencies among the vertices in the \prob{Clique} instance are encoded by time stamps.
See \cref{fig:hardnessB}(a) for an illustration.
So suppose that a \prob{Clique} instance with 
the vertex set~$\{v_1,\ldots,v_n\}$,
edge set~$\{e_1,\ldots,e_m\}$, 
and solution size~$r$ is given.
The constructed underlying graph consists of two stars with~$m$ leaves each and adjacent centers~$x$ and~$y$.
The leaves of the first star are only connected to $x$ at time step 1, thus $x$ is the source vertex that reaches the most other vertices.
For each~$v_i$, the edge~$\{x,y\}$ is present at time step~$i\beta+2$.
For each edge~$e_\ell=\{v_i,v_j\}$,
the edge connecting~$y$ and the vertex~$e_\ell$ (the vertex corresponding to edge~$e_\ell$) is present at time steps~$i\beta+\alpha+2$ and~$j\beta+\alpha+2$.
Observe that reaching $e_\ell$ from $x$ by a strict $(\alpha, \beta)$-temporal path requires the edge $\{x, y\}$ to be present at time~$i\beta+2$ or at time~$j\beta+2$.
Thus, if~$v_{i_1},\ldots,v_{i_r}$ form a clique on~$r$ vertices,
then deleting the temporal edge set~$\{(\{x,y\},\ell\beta+2)\mid \ell\in\{i_1,\ldots,i_r\}\}$ reduces the $(\alpha,\beta)$-reachability of~$x$ by $\binom{r}{2}$.
Conversely, reducing the $(\alpha,\beta)$-reachability of~$x$ by $\binom{r}{2}$ with only $r$ deletions is impossible unless a clique of size $r$ exists in the input.

\paragraph{Reordering of Layers.}
\citet{EnrightMS18} 
proved the following hardness result for the \prob{Min-Max Reachability Temporal Ordering (Min-Max RTO)} problem:
given a temporal graph~$\TG=(V,\TE,\tau)$ and an integer~$k\in\N$,
decide whether there is a bijection~$\phi:\set{\tau}\to\set{\tau}$ such that the maximum reachability 
(that is, the maximum number of vertices any vertex can reach via a strict temporal path) 
in~$\TG'=(V,\{(e,\phi(t))\mid (e,t)\in\TE\},\tau)$ is at most~$k$.
Correspondingly, \prob{Max-Min RTO} is defined by exchanging ``maximum'' by ``minimum'' and ``at most'' by ``at least''.
\begin{theorem}[\cite{EnrightMS18}]
 \label{thm:mmrto}
 \prob{Min-Max Reachability Temporal Ordering} and \prob{Max-Min Reachability Temporal Ordering} are \NP-hard 
 even when the underlying graph is a tree obtained by connecting two stars using a path.
\end{theorem}
\citeauthor{EnrightMS18} proved \cref{thm:mmrto} 
(similarly to the previously presented reduction by~\citet{EnrightMMZ18}) 
via a reduction from \prob{Clique}.
See \cref{fig:hardnessB}(b) for an illustration.
In their reduction,
the input consists of vertex set~$\{v_1,\ldots,v_n\}$,
edge set~$\{e_1,\ldots,e_m\}$, 
and solution size~$r$.
Each layer~$G_i$ corresponds to a vertex~$v_i$: 
the edge~$\{y,e_\ell\}$ is present in~$G_i$ if and only if~$v_i\in e_\ell$.
That is,
the incidence of an edge with vertex~$v_i$ is represented by the presence of that edge in layer~$G_i$.
Hence, 
if $v_{i_1},\ldots,v_{i_r}$ form a clique on~$r$ vertices,
then mapping~$i_1,\ldots,i_r$ to the first~$r$ layers disallows~$s$ to reach~$\binom{r}{2}$ leaves adjacent to~$y$ 
(since the $s$-$y$~path contains~$r+1$ vertices).

\paragraph{Merging of Layers.}

\citet{DeligkasmP19}
studied reachability minimization/maximization under certain layer-merging operations and showed hardness results on trees and paths.
In this context, 
\emph{merging} an interval of time~stamps $M \subseteq \set{\tau}$ in $\TG$ means replacing each temporal edge $(e, \ell)$ with $\ell \in M$ by a new temporal edge $(e, \max(M))$.
Thus, 
the appearance of all temporal edges within this interval~$M$ is shifted to the end of~$M$.
More precisely,
\citeauthor{DeligkasmP19} considered the \prob{Min Reachability Temporal Merging (MRTM)} problem:
given a temporal graph~$\TG=(V,\TE,\tau)$, 
a set of sources $S \subseteq V$, 
and three integers~$\lambda, \mu, k \in\N$,
decide  whether there are~$\mu$ disjoint intervals~$M_1, \dots, M_\mu$, 
each of size $|M_i \cap \set{\tau}| = \lambda$,
such that, 
after merging each of them in $\TG$, 
the number of vertices reachable from $S$ is at most $k$.
\begin{theorem}[\cite{DeligkasmP19}]
 \label{thm:mrtm}
 \prob{Min Reachability Temporal Merging} is \NP-hard 
 even when the underlying graph is a path.
\end{theorem}

The proof of \cref{thm:mrtm} employs a reduction from \prob{Max2SAT(3)},
a variant of the \prob{Max2SAT} problem 
where each variable occurs in at most three clauses.
(In the \prob{Max2SAT} problem, the goal is to find a truth assignment maximizing the number of satisfied clauses of a given 2-SAT formula.)

\begin{figure*}[t!]
 \centering
 \begin{tikzpicture}
 
  \tikzstyle{xnode}=[circle, fill=black, draw, scale=0.6]
  \tikzstyle{xedge}=[thick,-]

  \def\xr{0.9}
  \def\yr{0.9}
  
  \def\lvdist{1}
  \def\lhdist{0.4}
  \newcommand{\astar}[5]{
    \node (#3center) at (#1,#2)[xnode]{};
    \node (#3l1) at (#1-3*\lhdist*\xr,#2+#4*\lvdist*\yr)[xnode]{};
    \node (#3l2) at (#1-\lhdist*\xr,#2+#4*\lvdist*\yr)[xnode]{};
    \node (#3li) at (#1+\lhdist*\xr,#2+#4*\lvdist*\yr)[]{#5};
    \node (#3ln) at (#1+3*\lhdist*\xr,#2+#4*\lvdist*\yr)[xnode]{};
    \foreach \x in {1,2,i,n}{\draw[xedge] (#3center) to (#3l\x);}
  }
  
  \begin{scope}[xshift=\xr*0cm,yshift=-\yr*5.5cm]
    \node at (-2*\xr,0.75*\yr)[]{(a)};
    \node (root) at (0,0)[xnode,label=above:$s$]{};
    \foreach \y in {1,2,...,4}{
		\node[xnode] (l\y) at (-0.5*\xr,-0.7*\y*\yr){};
		\node[xnode] (r\y) at (0.5*\xr, -0.7*\y*\yr){};
	}
    \draw[xedge, every node/.style={midway,left}] 
		(root) -- (l1) node {$4c$}
		       -- (l2) node {$4c+1$}
		       -- (l3) node {$4i$}
		       -- (l4) node {$4i+1$};
	\draw[xedge, every node/.style={midway,right}]
		(root) -- (r1) node {$4c+1$}
		       -- (r2) node {$4c+2$}
		       -- (r3) node {$4j+1$}
		       -- (r4) node {$4j+2$};
	\draw[xedge]
		(l4) -- node[label=left:\ldots] {} (-1*\xr,-0.7*4*\yr);
	\draw[xedge]
		(r4) -- node[label=right:\ldots] {} (1*\xr,-0.7*4*\yr);
  \end{scope}
  
  \begin{scope}[xshift=\xr*8.5cm,yshift=-\yr*6.25cm]
  
    \node at (-5*\xr,1.5*\yr)[]{(b)};
    
    \foreach \x in {0,1,2}{
    \draw[gray,thin,densely dotted] (0*\xr,1*\yr-\x*\yr) -- (1.5*\xr,1*\yr-\x*\yr);
    \draw[gray,thin,densely dotted] (0.75*\x*\xr,1*\yr) -- (0.75*\x*\xr,-1*\yr);
    }
    \node (11) at (0*\xr,1*\yr)[scale=0.6,label=180:{$e_1$}]{};
    \node (12) at (0.75*\xr,1*\yr)[xnode]{};
    \node (13) at (1.5*\xr,1*\yr)[xnode]{};
    \node (21) at (0*\xr,0*\yr)[xnode,label=180:{$e_2$}]{};
    \node (31) at (0*\xr,-1*\yr)[xnode,label=180:{$e_3$},label=-90:{$1$}]{};
    \node (32) at (0.75*\xr,-1*\yr)[xnode,label=-90:{$2$}]{};
    \node (33) at (1.5*\xr,-1*\yr)[scale=0.6,label=-90:{$3$}]{};
    \draw[xedge] (32) -- (31) -- (21) -- (12) -- (13);
    \draw[xedge] (32) -- (21);
    
    \node at (-1.5*\xr,0)[]{$\leftrightarrow$};
    
    \foreach \x in {0,1,...,3}{\node (v\x) at (-3.5*\xr,1*\yr-0.75*\x*\xr)[xnode]{};}
    \draw[xedge] (v0) to (v3);
    \node[below left=of v0,xshift=\xr*3em,yshift=\yr*3.25em] (a){$e_1$};
    \node[below right=of v0,xshift=-\xr*3em,yshift=\yr*3.5em] (a){$(2,3)$};
    \node[below left=of v1,xshift=\xr*3em,yshift=\yr*3.25em] (a){$e_2$};
    \node[below right=of v1,xshift=-\xr*3em,yshift=\yr*3.5em] (a){$1$};
    \node[below left=of v2,xshift=\xr*3em,yshift=\yr*3.25em] (a){$e_3$};
    \node[below right=of v2,xshift=-\xr*3em,yshift=\yr*3.5em] (a){$(1,2)$};
  \end{scope}
  \end{tikzpicture}
 \caption{Illustration of the reductions behind~\cref{thm:mrtm,thm:tm}. 
 (a) A part of the temporal graph used in the proof of \cref{thm:mrtm} whose underlying graph is a path~\cite{DeligkasmP19}.
 (b) A temporal graph whose underlying graph is a path (left-hand side) and its 2-temporal line graph (right-hand side; a grid is indicated by thin gray dotted lines)~\cite{MertziosMNZZ20}.
 }
 \label{fig:hardnessC}
\end{figure*}
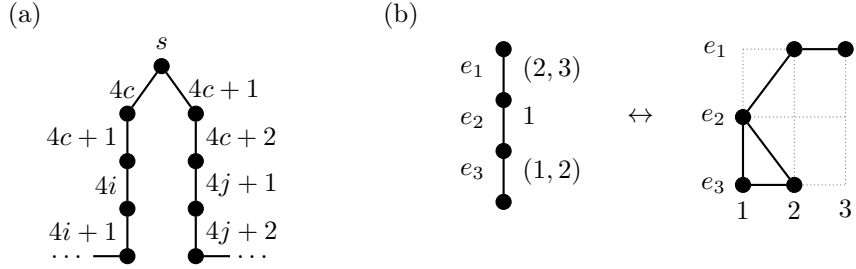
For each clause in a given input instance,
a separate subpath containing nine vertices of the underlying path is used and labeled as shown in \cref{fig:hardnessC}(a).
Here, 
$c$~is the index of the clause~$(x_i \lor \overline{x_j})$ and we may assume~$c$ to always be much smaller than~$i$ and~$j$.
The middle vertex $s$ of this subpath is added to the set $S$ of sources and we take the merge size as~$\lambda = 2$.
Then it is possible to either merge $\{4c, 4c+1\}$, 
thus preventing $s$ from reaching the three bottom left vertices, or to merge~$\{4c+1, 4c+2\}$, 
thus blocking the three bottom right vertices, 
but not both (due to the disjointness condition).
Hence, 
given a large enough number of merges, %
each source~$s$ can reach at most five other vertices.
If we want to reduce this number to four, 
then one must additionally merge $\{4i, 4i+1\}$ (thus setting $x_i$ to true) or merge~$\{4j+1, 4j+2\}$ (thus setting~$x_j$ to false), i.e., give an assignment satisfying clause~$c$.

If the underlying graph is allowed to be a ternary tree, then this construction can be modified to only require a single source vertex \cite{DeligkasmP19}.

\subsection{Temporal Matching}
\label{ssec:hardnessTM}
\citet{MertziosMNZZ20} 
proved hardness for the \prob{Temporal Matching (TM)} problem:
given a temporal graph~$\TG=(V,\TE,\tau)$ and integers~$k,\Delta\geq 0$,
decide whether there is a $\Delta$-temporal matching
of cardinality at least~$k$ in~$\TG$. A~\emph{$\Delta$-temporal matching} is a set~$\TE'\subseteq \TE$ of temporal edges such that for every two temporal~edges~$(e,t),(e',t')\in \TE'$,
we have that~$e\cap e'=\emptyset$ or~$|t-t'|\geq \Delta$.
\begin{theorem}[\cite{MertziosMNZZ20}]
 \label{thm:tm}
 \prob{Temporal Matching}
 is \NP-hard even when the underlying graph is a path.
\end{theorem}
The crucial observation is that solving \prob{TM} on a temporal graph $\TG$ is equivalent to solving \prob{Independent Set} on the so-called \emph{$\Delta$-temporal line graph} of $\TG$,
which contains a vertex for each temporal edge of $\TG$ and has two vertices adjacent if the corresponding temporal edges cannot be both contained in a $\Delta$-temporal matching~\cite[Definition~2]{MertziosMNZZ20}.
For an illustration see \cref{fig:hardnessC}(b).
Moreover, if the underlying graph $\ug{\TG}$ is a path with $m$ edges, then its $2$-temporal line graph is an induced subgraph of a \emph{diagonal grid graph} of size $m \times \tau$,
and conversely each such grid can be obtained as a $2$-temporal line graph.
Here, a \emph{diagonal grid graph} is simply a grid that additionally contains the two diagonal edges of every grid cell.
Subsequently, \citeauthor{MertziosMNZZ20} proved that \prob{Independent Set} is \NP-complete on induced subgraphs of diagonal grid graphs, 
thus also showing \NP-hardness of~\prob{TM}.

\section{Dynamic Programming Based on an Underlying Tree Decomposition}
\label{sec:tract}

\casablanca{It's still the same old story...}
For many graph problems,
algorithms exploiting small treewidth  
are dynamic programs over a corresponding tree decomposition. 
For temporal graph problems,
few such dynamic programs are known.
Yet,
we present four dynamic programs known from the literature:
Two \XP-algorithms for two \NP-hard problems,
and two polynomial-time algorithms.
For the former two algorithms,
the running time depends exponentially on the lifetime~$\tau$
(hence proving \fpty{} regarding~$\utw+\tau$ in both cases).
This supports our intuition that while capturing the structure of the graph,
the underlying treewidth is missing relevant time aspects.

\subsection{Two \XP-Algorithms}

The two \XP-algorithms~\cite{FluschnikMNRZ20,AxiotisF16} %
we sketch indeed
both are \FPT-algorithms regarding the combination~$\utw+\tau$ of the underlying treewidth and the lifetime.

\paragraph{An XP-Algorithm for Temporal Separation.}

\citet{FluschnikMNRZ20} studied \prob{Temporal Separation},
which is the problem of deciding whether all (strict) temporal paths connecting 
two given terminal vertices~$s$ and~$z$ in a temporal graph~$\TG$
can be destroyed by removing a set $S \subseteq V \setminus \{s, z\}$ of at most $k$ vertices.
Such a set $S$ is called a \emph{(strict) $s$-$z$~separator}.
\citet{FluschnikMNRZ20} employed dynamic programming on a given tree decomposition to prove 
that this problem is \fpt{} when parameterized by~$\utw+\tau$,
and is in~$\XP$ when parameterized by~$\utw$.
\begin{theorem}[\cite{FluschnikMNRZ20}]
	\label{thm:fpt-tw-tau}
	\prob{Temporal Separation} with given tree decomposition of the underlying graph is solvable in $O((\tau+2)^{\utw+2} \cdot \utw \cdot |V| \cdot |\TE|)$ time.
\end{theorem}

\begin{figure*}[t!]
	\includegraphics[width=\textwidth]{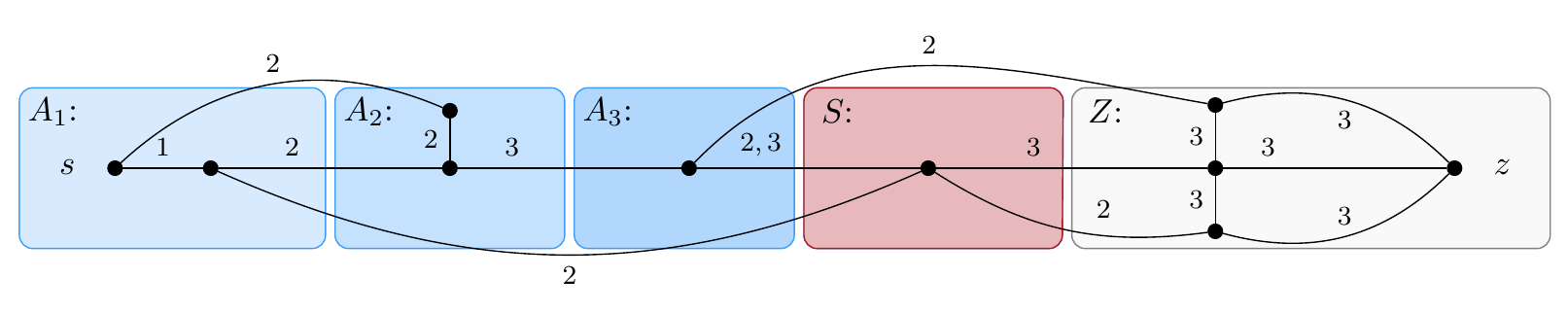}
	\caption{
		The idea for the dynamic program from \cref{thm:fpt-tw-tau} for a temporal graph~$\TG$. 
		Vertices in $S$ form the temporal $s$-$z$~separator,
		vertices in $Z$ are not reachable from $s$ in $\TG - S$, and
		vertices in $A_t$ are not reachable from $s$ in $\TG - S$ before time~$t$.
		}
	\label{fig:dp-idea}
\end{figure*}

\noindent 
The dynamic program behind \cref{thm:fpt-tw-tau} 
is based on the fact that for each vertex~$v \in V\setminus\{s\}$ in a temporal graph $\TG=(V,\TE,\tau)$ 
there is a time step~$t \in \set{\tau}$ such that~$v$ cannot be reached from~$s \in V$ before~$t$.
In particular, 
one guesses a partition~$V = A_1 \uplus A_2 \uplus \ldots \uplus A_\tau \uplus S \uplus Z$ 
such that 
(i)~$S$ is a temporal $s$-$z$~separator, 
(ii) in~$\TG-S$~no vertex contained in~$Z$ is reachable from~$s$, 
and (iii) no vertex~$v \in A_t$ can be reached from~$s$ before time step~$t$, where~$t \in \set{\tau}$.
See \cref{fig:dp-idea} for an illustrative example.

\paragraph{An \XP-Algorithm for Temporally Connected Subgraphs.}

Given a temporal graph $\TG=(V,\TE,\tau)$ and a designated vertex~$r\in V$,
a \emph{temporally $r$-connected spanning subgraph} of~$\TG$ is a temporal graph~$\TG'=(V,\TE',\tau)$ with~$\TE'\subseteq \TE$ such that
$\TG'$ contains a temporal path from~$r$ to any other vertex $v\in V$ in~$\TG'$.
The task in \prob{Minimum Single-Source Temporal Connectivity ($r$-MTC)} is to find a temporally $r$-connected spanning subgraph for
a given vertex $r$ such that the total weight~$\sum_{(e,t)\in \TE'} w((e,t))$ is minimized, where $w$ is an arbitrary nonnegative weight function.
\citet{AxiotisF16} employed dynamic programming on a nice tree decomposition~\cite{Kloks94}
to prove that \prob{Minimum Single-Source Temporal Connectivity ($r$-MTC)} is \fpt{} when parameterized by~$\utw+\tau$,
and is contained in~\XP{} when parameterized by~$\utw$ alone.

\begin{theorem}[\cite{AxiotisF16}]
 \label{thm:axofota}
 \prob{Minimum Single-Source Temporal Connectivity} with given nice tree decomposition of the underlying graph is solvable in~$\bigO(3^\utw \cdot (\tau+\utw)^{\utw+1}\cdot |V|)$ time. 
\end{theorem}

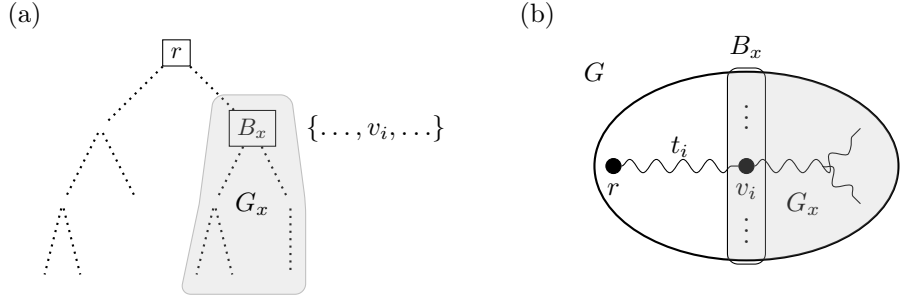
\begin{figure*}[t]
 \centering
 \begin{tikzpicture}

    \usetikzlibrary{calc,decorations.pathmorphing}
      \tikzstyle{tnode}=[rectangle,scale=0.9,draw]
      \tikzstyle{snode}=[scale=0.5]
      \tikzstyle{pnode}=[circle,fill,scale=0.6,draw]
      \tikzstyle{tedge}=[dotted,thick,-]
    
    \def\xr{1}
    \def\yr{1}
    \def\ep{0.2}

    \begin{scope}
    \node (r) at (-2*\xr,2.5*\yr)[]{(a)};
    \node (r) at (0*\xr,2*\yr)[tnode]{$r$};
    \node (lr) at (-1*\xr,1*\yr)[snode]{};
    \node (rr) at (1*\xr,1*\yr)[tnode,label=0:{$\phantom{=}\{\ldots,v_i,\ldots\}$}]{$B_x$};
    \node (llr) at (-1.5*\xr,0*\yr)[snode]{};
    \node (rlr) at (-0.5*\xr,0*\yr)[snode]{};
    \node (rrr) at (1.5*\xr,0*\yr)[snode]{};
    \node (lrr) at (0.5*\xr,0*\yr)[snode]{};

    \node (lllr) at (-1.75*\xr,-1*\yr)[snode]{};
    \node (rllr) at (-1.25*\xr,-1*\yr)[snode]{};

    \node (llrr) at (0.25*\xr,-1*\yr)[snode]{};
    \node (rlrr) at (0.75*\xr,-1*\yr)[snode]{};

    \node (crrr) at (1.5*\xr,-1*\yr)[snode]{};

    \draw[tedge] (r) to (lr);
    \draw[tedge] (r) to (rr);
    \draw[tedge] (lr) to (llr);
    \draw[tedge] (lr) to (rlr);
    \draw[tedge] (rr) to (lrr);
    \draw[tedge] (rr) to (rrr);

    \draw[tedge] (llr) to (lllr);
    \draw[tedge] (llr) to (rllr);

    \draw[tedge] (lrr) to (rlrr);
    \draw[tedge] (lrr) to (llrr);
    \draw[tedge] (rrr) to (crrr);

    \draw[-,thin,fill=lightgray,rounded corners,opacity=0.25] 
      ($(lrr)+(-\ep,0)$) to ($(rr.north west)+(-\ep,\ep)$) to ($(rr.north east)+(\ep,\ep)$) to ($(rrr)+(\ep,0)$) to ($(crrr)+(\ep,-\ep)$) to ($(llrr)+(-\ep,-\ep)$) to ($(lrr)+(-\ep,0)$);
    \node at (1.0*\xr, 0*\yr)[]{$G_x$};
    \end{scope}

    \begin{scope}[xshift=\xr*7.5 cm,yshift=\yr*0.5 cm]
    \node (r) at (-2.75*\xr,2.0*\yr)[]{(b)};
    \draw[thick] ellipse (\xr*2 and \yr*1.25);
    \node at (-2*\xr,1.25*\yr)[]{$G$};
    \node at (0,0)[rounded corners, rectangle, draw,minimum height=\yr*2.6 cm,minimum width=\xr*0.5 cm,label=90:{$B_x$}]{};
    \node at (0,0.75*\yr)[]{$\vdots$};
    \node (v) at (0,0.0*\yr)[pnode,label=-90:{$v_i$}]{};
    \node at (0,-0.75*\yr)[]{$\vdots$};
      \node (r) at (-1.75*\xr,0.0*\yr)[pnode,label=-90:{$r$}]{};
      
    \draw[decorate,decoration=snake,-] (r) to node[midway,above]{$t_i$}(v);
    \draw[decorate,decoration=snake,-] (v) to ($(v)+(1*\xr,0*\yr)$) to ($(v)+(1.5*\xr,0.5*\yr)$);
    \draw[decorate,decoration=snake,-] ($(v)+(1*\xr,0*\yr)$) to ($(v)+(1.5*\xr,-0.5*\yr)$);
    \node at (0.75*\xr,-0.5*\yr)[]{$G_x$};
      \clip ellipse (\xr*2 and \yr*1.25);
      \draw[fill=lightgray,opacity=0.25] (-0.25*\xr,1.25*\yr) rectangle (2*\xr,-1.25*\yr);
    \end{scope}
  \end{tikzpicture}
  \caption{Illustration to the dynamic program for~\prob{$r$-MTC} for some temporal graph with underlying graph~$G$.
  (a) A (nice) tree decomposition of~$G$ is depicted with the root node's bag containing~$r$,
  node~$x$ with bag~$B_x=\{\dots,v_i,\dots\}$, 
  and the subgraph~$G_x$ of $G$ that is induced by all vertices contained in the bag~$B_x$ and bags of the descendants of~$x$.
  (b) Graph~$G$ is depicted, with induced subgraph~$G_x$.
  Moreover, a temporal $r$-$v_i$~path arriving at time step~$t_i$,
  and temporal paths connecting~$v_i$ to some vertices from~$G_x$ are depicted.
  The latter corresponds to an table entry~$f(x|\dots,a_i=1,t_i,\dots)$.}
  \label{fig:rmtc}
\end{figure*}

\noindent
The idea of the dynamic program behind~\cref{thm:axofota} using a nice tree decomposition rooted at the source~$r$ is as follows
(see \cref{fig:rmtc} for an illustration):
for each bag,
the vertices contained in the bag are (bi-)partitioned into vertices connected to~$r$ (we call them \emph{local sources}) and
vertices not (yet) connected to~$r$.
Vertices in the subgraph induced by the vertices in the bag and all its descendant bags must be reachable from the local sources
by a temporal path starting ``late enough'' 
(i.e., after the local source has been reached from source~$r$).
For each node~$x$ in the nice tree decomposition,
a table entry~$f(x\mid a_1,t_1,\dots,a_\utw,t_\utw)$
stores the minimum cost of a temporal subgraph such that in the graph induced by all the vertices in the bags of~$x$ and all its descendants,
every vertex is reachable from some vertex~$v^x_j\in B_x$ with~$a_j=1$ (the local sources) by some temporal path starting at time step~$t_j$ or later.
Hence, each such node $x$ has a table entry for each of the $2^{\utw}$~possible bipartitions,
and each of the~$\tau^{\utw}$ possible starting times for the temporal paths starting at local sources.

For both of the two presented problems,
it appears to be crucial to guess the time steps in which a solution ``touches'' the corresponding bag.
However,
in both it is open whether the dependencies on~$\tau$ 
(being the base of exponent~$\utw$; see~\cref{thm:fpt-tw-tau,thm:axofota})
can be avoided:
 Is \prob{Temporal Separation (TS)} or \prob{Minimum Single-Source Temporal Connectivity ($r$-MTC)} \fpt{} when parameterized by~$\utw$?

\subsection{Two Fixed-Parameter Polynomial-Time Algorithms}

The underlying tree decomposition,
when part of the input,
can also be used for tasks solvable in polynomial time.
In this section,
we present two algorithms making use of the underlying tree decomposition,
one for temporal exploration,
and one for computing foremost temporal walks.

\paragraph{Temporal Exploration.}

In~\cref{sec:intrac},
we discussed the \NP-hardness of determining the exact time required to fully explore a temporal graph.
However, as long as each layer of the input temporal graph is connected and the underlying treewidth is low,
it can be shown that a subquadratic number of steps is always sufficient.
More precisely, 
\citet{Erlebach0K15} proved the following by giving an algorithm that utilizes a given tree decomposition.

\begin{theorem}[\cite{Erlebach0K15}]\label{thm:exploration_tw}
	If every layer is connected, then a temporal graph $\TG$ can be explored in $\bigO\left(|V|^{3/2}\cdot\utw^{3/2}\cdot\log |V|\right)$ steps.
\end{theorem}

The proof of \cref{thm:exploration_tw} builds upon the observation that an agent needs at most $n-1$ steps to move from any vertex to any other vertex if both of these vertices are connected in every layer.
The idea is then to divide up $\ug{G}(\TG)$ into sufficiently small subgraphs to which this observation can then be applied.

To this end, 
select a vertex set~$S$ as the union of $\bigO(\sqrt{|V|\cdot\utw{}})$ bags of a nice tree decomposition of $\ug{G}(\TG)$
in such a way that every connected component of $\ug{G}(\TG) - S$ has size at most $\bigO(\sqrt{|V|/\utw{}})$.
If we consider a time window of $\Theta(\utw{} \cdot\sqrt{|V|/\utw{}})$ layers, then, by the pigeonhole principle, 
for any vertex $v$ in any of these connected components, there is a vertex $w \in S$ that is in the same connected component in at least $\Theta(\sqrt{|V|/\utw{}})$ layers.
Thus, 
by the above observation, 
an agent at~$w$ can reach~$v$ and return to~$w$ within $\bigO(\utw{}\cdot \sqrt{|V|/\utw{}})$ time steps.

Hence, 
if we use $\Theta(\utw{}\cdot \sqrt{|V|\cdot\utw{}})$ agents, 
then each starting at a vertex of~$S$, we can explore $\TG$ in at most $\bigO(\utw{}\cdot \sqrt{|V|/\utw{}}  \cdot \sqrt{|V|/\utw{}}) = \bigO(|V|)$ steps.
From this, 
one can derive an upper~bound of~$\bigO(|V|^{3/2}\cdot \utw{}^{3/2} \cdot\tlog{|V|})$ steps if only a single agent is used to perform these explorations sequentially.

\paragraph{Computing Foremost Walks.}

A (strict) \emph{foremost} $s$-$z$ walk is a temporal walk that arrives earliest among all~$s$-$z$ temporal walks.
\citet{Himmel2018} proved that foremost walk queries can be answered quickly
using a specific data structure that relies on a (given) underlying tree decomposition.

\begin{theorem}[\cite{Himmel2018}]
 \label{thm:himmel}
 There exists a data structure of size~$\bigO(\utw^2\cdot\tau\cdot |V|)$ computable in~$\bigO(\utw^2\cdot\tau^2\cdot |V|)$ time such that
 one can find a foremost walk between two vertices on 
 temporal graphs with underlying treewidth~$\utw$
 in~$\bigO(\utw^2\cdot\tau \cdot\log{|V|}\cdot\tlog{\utw\cdot\tau\cdot\log{|V|}})$ time.
\end{theorem}

The data structure behind~\cref{thm:himmel}
was originally introduced by~\citet{AbrahamCDGW16} for computing shortest path queries in static graphs.
It exploits \emph{binary} tree decompositions of depth~$\bigO(\log{|V|})$.
Basically,
the preprocessing for the data structure computes the earliest arrival time from any vertex~$v$ to any vertex~$w$ at any possible starting time,
where~$v$ and~$w$ are contained in the same bag of the tree decomposition.

\section{Monadic Second-Order Logic for Temporal Graphs}
\label{sec:mso}
\casablanca{Honest as the day is long!}
Courcelle's famous theorem
states that every graph property definable in monadic second-order logic 
is fixed-parameter tractable when simultaneously parameterized
by the treewidth of the graph and the length of the formula \cite{courcelle2012graph}.
In this section, we review how one could lift this powerful classification tool to temporal graphs
and spot some pitfalls having led to flaws in the literature.

For \emph{monadic second-order (\MSO) logic} on a graph $G$
we need a \emph{structure} consisting of
a universe $U = V(G) \cup E(G)$ and a \emph{vocabulary} consisting of 
two unary relations~$V \subseteq U$ and~$E\subseteq U$ containing the vertices and the edges,
respectively,
and two binary relations~${\rm adj} \subseteq U\times U$ and~${\rm inc}\subseteq U\times U$, where
$(v,w)\in {\rm adj}$ if and only if $\{v,w\}\in E(G)$,
and $(v,e)\in {\rm inc}$ if and only if~$v\in e$.
For a fixed finite set of \emph{(monadic) variables},
an \emph{atomic formula} over vocabulary~$\nu$ 
is of the form $x_1=x_2$ or $R(x_1, x_2)$ or $R'(x_1)$,
where $R \in \{{\rm adj},{ \rm inc}\}$, $R' \in \{V,E\}$ and $x_1,x_2 \in U$.
Here, $R(x_1,x_2)$ ($R'(x_1)$) evaluates to true if and only if $(x_1,x_2) \in R$ ($x_1\in R'$).
\emph{\MSO{} formulas} 
are constructed from atomic formulas 
using boolean operations $\neg$, $\vee$, $\wedge$ 
and existential and universal quantifiers $\exists$, $\forall$
over variables and set variables.
For further details, refer to \citet{courcelle2012graph}.
\begin{example}
\label{example:mso}
The well-known \prob{Clique} problem can be expressed by the following \MSO{} 
formula: $\exists X. (\forall x,y \in X. (V(x) \wedge V(y) \wedge {\rm adj}(x,y))).$
\end{example}
The following,
known as an optimization variant of Courcelle's theorem,
connects \MSO{} and treewidth.
\begin{theorem}[\cite{ARNBORG1991308,courcelle2012graph}]
	\label{thm:mso-opt}
	There exists an algorithm that, given
	\begin{inparaenum}[(i)]
		\item an \MSO{} formula $\rho$ with free monadic variables $X_1,\dots,X_r$,
		\item an $n$-vertex graph $G$, 
		and 
		\item an affine function $\alpha(x_1,\dots, x_r)$,
	\end{inparaenum}
	finds the minimum (maximum) of $\alpha(|X_1|,\dots,|X_r|)$ 
	over evaluations of~$X_1,\dots,X_r$ for which formula~$\rho$ is satisfied on $G$.
	The running time of this algorithm is~$f(|\rho|,\tw(G)) \cdot n$, 
	where $f$ is a computable function,
	$|\rho|$ is the length of~$\rho$,
	and~$\tw(G)$ is the treewidth of~$G$.
\end{theorem}

\noindent
Having \cref{thm:mso-opt} at hand,
we can prove that \prob{Clique} (see~\cref{example:mso})
is fixed-parameter tractable when parameterized by the treewidth of the input graph:
The formula given in~\cref{example:mso} has one free monadic variable and constant length~$c$,
hence,
with $\alpha$~being the identity function,
we can decide \prob{Clique} in~$f(c,\tw(G))\cdot |V|$ time.

We are aware of two successful approaches 
and one flawed approach
to lift 
\cref{thm:mso-opt}
to the temporal setting. 
We will survey in \cref{ssec:usinglabels,ssec:enrichingvoc} the two successful approaches,
and discuss in \cref{ssec:pitfall} the flawed approach.

\subsection{Using Labels}
\label{ssec:usinglabels}

\citet{ARNBORG1991308} showed that it is possible to apply
\cref{thm:mso-opt} to graphs in which edges have labels from a fixed finite set,
either by augmenting the graph logic to incorporate predicates describing the labels,
or by representing the labels by unquantified edge set variables.
\citet{ZschocheFMN18} exploited this for temporal graphs as follows 
(see~\cref{fig:tempgrlabelgraph} for an example):
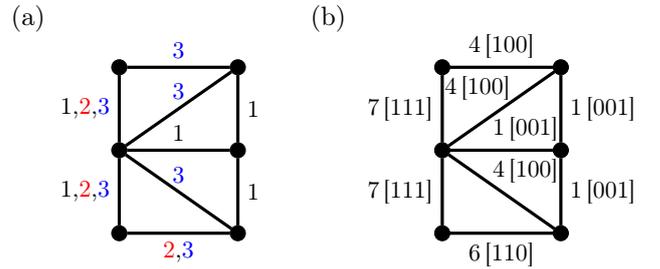
\begin{figure}[t]
 \centering
 \begin{tikzpicture}

  \usetikzlibrary{calc}
  \tikzstyle{xnode}=[fill,circle,draw,scale=0.6];
  \tikzstyle{xedge}=[very thick,-];
  \tikzstyle{xlabel}=[midway,scale=0.9];

  \def\xr{1}
  \def\yr{1}
  \def\xsc{1.3}
  \def\ysc{1.1}

  \newcommand{\hlbnodes}[2]{
  \node (l2) at (#1,#2+1*\ysc*\yr)[xnode]{};
    \node (l1) at (#1,#2)[xnode]{};
  \node (l0) at (#1,#2-1*\ysc*\yr)[xnode]{};
  \node (r2) at (#1+1*\xsc*\xr,#2+1*\ysc*\yr)[xnode]{};
    \node (r1) at (#1+1*\xsc*\xr,#2)[xnode]{};
  \node (r0) at (#1+1*\xsc*\xr,#2-1*\ysc*\yr)[xnode]{};
  }

  \begin{scope}[xshift=\xr*0.25 cm,xscale=1.2]
  \node at (-1.0*\xr,1.75*\yr)[]{(a)};
    \hlbnodes{0}{0};
    \draw[xedge] (l1) to node[xlabel,left]{1,\textcolor{red}{2},\textcolor{blue}{3}}(l0);
    \draw[xedge] (l1) to node[xlabel,left]{1,\textcolor{red}{2},\textcolor{blue}{3}}(l2);

    \draw[xedge] (r1) to node[xlabel,right]{1}(r0);
    \draw[xedge] (r1) to node[xlabel,right]{1}(r2);

    \draw[xedge] (l1) to node[xlabel,above]{1}(r1);
    \draw[xedge] (l0) to node[xlabel,below]{\textcolor{red}{2},\textcolor{blue}{3}}(r0);
    \draw[xedge] (l2) to node[xlabel,above]{\textcolor{blue}{3}}(r2);

    \draw[xedge] (l1) to node[xlabel,above]{\textcolor{blue}{3}}(r0);
    \draw[xedge] (l1) to node[xlabel,above]{\textcolor{blue}{3}}(r2);
  \end{scope}

  \def\xsh{5.5}
  \ifdoublecolumn{}\def\xsh{4.5}\fi{}
  \begin{scope}[xshift=\xr*\xsh cm,xscale=1.2]
  \node at (-1.25*\xr,1.75*\yr)[]{(b)};
    \hlbnodes{0}{0};
    \draw[xedge] (l1) to node[xlabel,left]{7\,[111]}(l0);
    \draw[xedge] (l1) to node[xlabel,left]{7\,[111]}(l2);

    \draw[xedge] (r1) to node[xlabel,right]{1\,[001]}(r0);
    \draw[xedge] (r1) to node[xlabel,right]{1\,[001]}(r2);

    \draw[xedge] (l1) to node[xlabel,above,xshift=\xr*1em]{1\,[001]}(r1);
    \draw[xedge] (l0) to node[xlabel,below]{6\,[110]}(r0);
    \draw[xedge] (l2) to node[xlabel,above]{4\,[100]}(r2);

    \draw[xedge] (l1) to node[xlabel,above,xshift=\xr*1em]{4\,[100]}(r0);
    \draw[xedge] (l1) to node[xlabel,above,xshift=-\xr*1em]{4\,[100]}(r2);
  \end{scope}
  \end{tikzpicture}
  \caption{(a) The temporal graph~$\TG$ from~\cref{fig:tempgr}  and (b) its edge-labeled graph~$L(\TG)$ 
  (with labels and bit-representation in brackets) 
  are depicted.}
  \label{fig:tempgrlabelgraph}
\end{figure}
For a given temporal graph $\TG$ of lifetime~$\tau$,
define the edge-labeled graph~$L(\TG)$ as the underlying graph~$\ug{G}$
with the added
edge-labeling $\omega \colon E(\ug{G}) \rightarrow \set{2^{\tau} -1}$ 
such that $\omega(\{v,w\}) = \sum^{\tau}_{i=1} \mathds{1}_{\{v,w\} \in E_i} \cdot 2^{i-1}$, 
where $\mathds{1}_{\{v,w\} \in E_i}=1$ 
if and only if~$(\{v,w\},i)\in\TE$, and $0$ otherwise.
Observe that the $i$-th bit of a label now expresses whether the edge is present 
in the $i$-th layer of the temporal graph.
Hence, we can check whether an edge $e$ is present in layer~$t$
using the \MSO{} formula
$
{\rm layer}(e,t) :=~\bigvee_{i=1}^\tau \bigvee_{j\in \sigma(i, 2^\tau-1)} \big(t = i~\land \omega(e) = j\big) 
$
of length $2^{O(\tau)}$, where $\sigma(i,z) := \{ x \in \set{z} \mid i\text{-th bit}$ $\text{of }x\text{ is 1} \}$.
Furthermore, we can determine whether 
two vertices~$v$ and~$w$ are adjacent in layer~$t$
using the \MSO{} formula 
$
{\rm tadj}(v,w,t) :=~\exists e \in E. \big({\rm inc}(e,v) \land {\rm inc}(e,w) \land {\rm layer}(e,t)\big)
$
of length~$2^{O(\tau)}$.
Altogether, 
in a nutshell we get the following:
  If a temporal graph problem~$\Pi$
  can be formulated
  by an \MSO{}-formula which 
  uses~${\rm layer}(e,t)$ and~${\rm tadj}(v,w,t)$ 
  as black boxes,
  then 
  $\Pi$ is \fpt{} when
  parameterized by the combination of 
  the length of the formula, 
  the underlying treewidth, 
  and the lifetime $\tau$.
\citet{ZschocheFMN18} derived an \MSO{}-formula for 
\prob{Temporal Separation~(TS)},
where the length of the formula is upper-bounded by some function in~$\tau$.
Hence, 
\prob{TS} is \fpt{}
when parameterized by 
the combination of 
the underlying treewidth
and the lifetime.

\subsection{Enriching the Vocabulary}
\label{ssec:enrichingvoc}

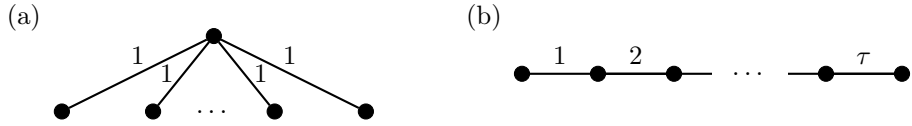
\begin{figure*}[t]
	\begin{center}
\begin{tikzpicture}
  	\tikzstyle{xnode}=[circle, fill=black, draw, scale=0.6]
  	\tikzstyle{xedge}=[thick,-]
  	\tikzstyle{yedge}=[thick,-]
  	\tikzstyle{zedge}=[thick,-,dotted]

	\begin{scope}
	 
	\node at (-2.5, 0.25)  {(a)};
  \node[xnode] at (0, 0) (r)  {};
	\node[xnode] at (-2, -1) (l1)  {};
	\node[xnode] at (-0.8, -1) (l2)  {};
	\node at (0, -1) (ldots)  {$\dots$};
	\node[xnode] at (0.8, -1) (l3)  {};
	\node[xnode] at (2, -1) (l4)  {};

	\draw[yedge] (r) to node[above] {$1$} (l1); 
	\draw[yedge] (r) to node[left] {$1$} (l2); 
	\draw[yedge] (r) to node[right] {$1$} (l3); 
	\draw[yedge] (r) to node[above] {$1$} (l4); 
	\end{scope}

	\begin{scope}[xshift=3em]
	\node[xnode] at (3, -0.5) (p1)  {};
	\node[xnode] at (4, -0.5) (p2)  {};
	\node[xnode] at (5, -0.5) (p3)  {};
	\node[xnode] at (7, -0.5) (p4)  {};
	\node[xnode] at (8, -0.5) (p5)  {};

	\draw[yedge] (p1) to node[above] {$1$} (p2); 
	\draw[yedge] (p2) to node[above] {$2$} (p3); 
	\node at (6, -0.5)  {$\dots$};
	\draw[yedge] (p4) to node[above] {$\tau$} (p5); 
	\draw[yedge] (p2) to  (5.5,-0.5); 
	\draw[yedge] (6.5,-0.5) to (p5); 
	
	\node at (2.5, 0.25)  {(b)};
	\end{scope}
	
\end{tikzpicture}

	\end{center}
	\caption{An example why the parameters $\utw + \tau$ 
	and $\utw + \Delta_{\TG}$ are incomparable. 
	Both temporal graphs have constant treewidth of the underlying graph.
	The temporal graph in (a) has only one layer but
	the maximum temporal total degree is unbounded.
	The temporal graph in (b) has maximum temporal total degree of two but an unbounded number of layers.}

\label{fig:mttd-vs-tau}
\end{figure*}

Another approach, used by \citet{EnrightMMZ18},
can be applied to exchange the dependency on $\tau$ 
with a dependency on the \emph{maximum temporal total degree}~$\Delta_{\TG}$,
which is the maximum number of temporal edges incident to the same vertex in temporal graph~$\TG$.
Observe that the maximum temporal total degree is 
at least the maximum degree of the underlying graph.
Moreover, 
the parameters lifetime and maximum temporal total degree are unrelated to each other, 
meaning that the maximum temporal total degree can be large while the lifetime is small
and vice versa (see \cref{fig:mttd-vs-tau} for two examples). 

In a nutshell, 
we alter the universe and the vocabulary
of the structure (we refer to this structure as \emph{enriched})
in order to express a temporal graph problem.
We add all temporal edges~$(e,t)$ of the temporal graph~$\TG$ 
to the universe and equip the vocabulary with two 
binary relation symbols~$\mathcal L$ and~$\mathcal R$,
where
\begin{compactitem}
 \item $(e,(e,t)) \in \mathcal L$ if and only if~$e$ is an edge in the underlying graph 
and $(e,t)$~is a temporal edge of~$\TG$, and
 \item $(e_1,t_1),(e_2,t_2) \in \mathcal R$  if and only if~$(e_1,t_1)$ and~$(e_2,t_2)$ are temporal edges where~$e_1$ and $e_2$~have a vertex in common and $t_1 < t_2$.
\end{compactitem}
It is easy to see that the treewidth of 
the Gaifman graph\footnote{%
In the \emph{Gaifman graph} of a structure, there is one vertex 
for each element in the universe and two vertices have an edge
if and only if the corresponding elements occur together in the same relation.}
for the enriched structure is upper-bounded by
a function of the treewidth of the underlying graph of the temporal graph and
the maximum temporal total degree.
Hence, 
due to~\citet{courcelle2012graph}, 
if a temporal graph problem~$\Pi$
can be formulated by an \MSO{}-formula in the enriched structure,
then~$\Pi$ is \fpt{} when parameterized by the combination of the underlying treewidth,
the maximum temporal total degree, 
and the length of that formula.
\citet{EnrightMMZ18} derived an \MSO{}-formula 
in the enriched structure for 
\prob{Temporal Reachability Edge Deletion},
where the length of the formula depends on~$h$ 
(the size of the set of reachable vertices),
hence proving
\fpty{} for the problem
when parameterized by the combination of~$h$, 
the underlying treewidth, and
the maximum temporal total degree.

\subsection{Pitfalls in the Literature}
\label{ssec:pitfall}

\citet{MansM14} also explored
the direction of enriching the vocabulary 
in the context of \emph{dynamic graphs}.
In their model, vertices can (dis)appear over time as well.
Furthermore, the layers are not necessarily arranged in a linear (time) ordering. 
Hence, their model of dynamic graphs is more general than temporal graphs.
However, some of their results seem flawed.
In the remainder of this section we discuss 
these flaws in the special case of temporal graphs.

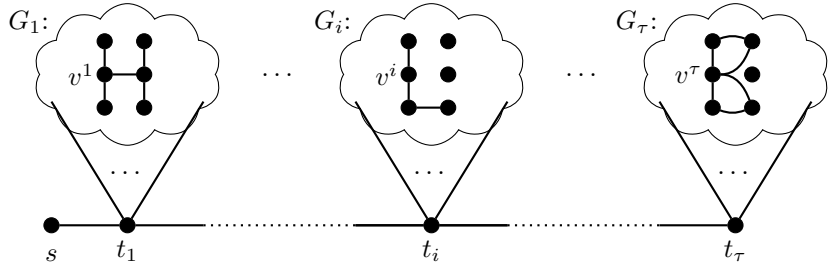
\begin{figure*}[t]
	\centering
\begin{tikzpicture}
  	\tikzstyle{xnode}=[circle, fill=black, draw, scale=0.6]
  	\tikzstyle{xedge}=[thick,-]
  	\tikzstyle{yedge}=[thick,-]
  	\tikzstyle{zedge}=[thick,-,dotted]
  	\tikzstyle{l1edge}=[xedge]
  	\tikzstyle{l2edge}=[xedge]
  	\tikzstyle{l3edge}=[xedge]
	
	\usetikzlibrary{calc}
  \tikzstyle{xlabel}=[midway,scale=0.9];

  \def\xr{1}
  \def\yr{1}
  \def\xsc{1.3}
  \def\ysc{1.1}

  \newcommand{\hlbnodes}[2]{
  \node (l2) at (#1,#2+1*\ysc*\yr)[xnode]{};
    \node (l1) at (#1,#2)[xnode]{};
  \node (l0) at (#1,#2-1*\ysc*\yr)[xnode]{};
  \node (r2) at (#1+1*\xsc*\xr,#2+1*\ysc*\yr)[xnode]{};
    \node (r1) at (#1+1*\xsc*\xr,#2)[xnode]{};
  \node (r0) at (#1+1*\xsc*\xr,#2-1*\ysc*\yr)[xnode]{};
  }
	
	\node[xnode,label={[shift={(0.0,-0.7)}]$s$}] at (-1, 0) (b)  {};
	\node[xnode,label={[shift={(0.0,-0.7)}]$t_1$}] at (0, 0) (t1)  {};
	\node[xnode,label={[shift={(0.0,-0.7)}]$t_i$}] at (4, 0) (ti)  {};
	\node[xnode,label={[shift={(0.0,-0.7)}]$t_\tau$}] at (8, 0) (tt)  {};

	\draw[yedge] (b) to (t1) to (1,0); 
	\draw[yedge] (3,0) to (ti) to (5,0); 
	\draw[yedge] (3,0) to (ti) to (5,0); 
	\draw[yedge] (7,0) to (tt); 
	\draw[zedge] (1,0) to (3,0); 
	\draw[zedge] (5,0) to (7,0); 

	\draw[yedge] (t1) to (-1,1.64); 
	\draw[yedge] (t1) to (1,1.64); 
	\node  at (0,0.7) {$\dots$};

	\draw[yedge] (ti) to (3,1.64); 
	\draw[yedge] (ti) to (5,1.64); 
	\node  at (4,0.7) {$\dots$};

	\draw[yedge] (tt) to (7,1.64); 
	\draw[yedge] (tt) to (9,1.64); 
	\node  at (8,0.7) {$\dots$};
	\node[cloud, draw,cloud puffs=10,cloud puff arc=120, aspect=1.5, inner ysep=1.55em] at (0,2) (G1)  {};
	\node at (-1.3,2.7) {$G_1$:};

	\node at (2,2) {$\dots$};

	\node[cloud, draw,cloud puffs=10,cloud puff arc=120, aspect=1.5, inner ysep=1.55em] at (4,2)  {};
	\node at (4-1.3,2.7) {$G_i$:};

	\node at (6,2) {$\dots$};

	\node[cloud, draw,cloud puffs=10,cloud puff arc=120, aspect=1.5, inner ysep=1.55em] at (8,2)  {};
	\node at (8-1.3,2.7) {$G_{\tau}$:};

	\begin{scope}[yshift=\yr*2.0cm,scale=0.4]
    \hlbnodes{-0.75*\xr}{0};
    \draw[l1edge] (l1) to (l0);
    \draw[l1edge] (l1) to (l2);
    \draw[l1edge] (r1) to (r0);
    \draw[l1edge] (r1) to (r2);
    \draw[l1edge] (l1) to (r1);
    \node[left=of l1,xshift=\xr*3.1em] (lb)[]{$v^1$};

    \hlbnodes{9.25*\xr}{0};
    \draw[l2edge] (l1) to (l0);
    \draw[l2edge] (l1) to (l2);
    \draw[l2edge] (l0) to (r0);
    \node[left=of l1,xshift=\xr*3.1em] (lb)[]{$v^i$};

    \hlbnodes{19.25*\xr}{0};
    \draw[l3edge] (l1) to (l0);
    \draw[l3edge] (l1) to (l2);
    \draw[l3edge] (l0) to [out=-20,in=-160](r0);
    \draw[l3edge] (l2) to [out=20,in=160](r2);
    \draw[l3edge] (l1) to [out=0,in=110](r0);
    \draw[l3edge] (l1) to [out=0,in=-110](r2);
    \node[left=of l1,xshift=\xr*3.1em] (lb)[]{$v^\tau$};
  \end{scope}
\end{tikzpicture}
\caption{Rough sketch of the Gaifman graph of a \tps{} for a temporal graph~$\TG$ with layers~$G_1,\ldots,G_\tau$. 
The copies~$v^1,v^i,v^\tau$ are illustrated for a vertex~$v$ of~$\TG$.}
\label{fig:wrongstuff}
\end{figure*}

\citeauthor{MansM14} construct a so-called \emph{\tps{}}.
Here, the universe 
has for each vertex~$v$ of the temporal graph $\tau$~many copies~$v^1,$ $\dots,v^{\tau}$, one element~$t_i$ for each~$i \in \set{\tau}$,
and an additional element $s$.
Note, that there is a unary relation symbol~$L_v$ which 
contains an element $x$ if and only if the element $x$ is generated from the vertex~$v$ ($x\equiv v^t$, for some $t \in \set{\tau}$).
The Gaifman graph of a \tps{} of a temporal graph is the disjoint union of the layers.
Additionally, there is one long path 
starting at some special vertex $s$ 
and then ``visits'' all layers in the time induced order, see \cref{fig:wrongstuff} for an illustration.

On the good side, this keeps the treewidth of the Gaifman graph upper-bounded by a function 
in the maximum treewidth over all layers.
Furthermore, one can still express (in \MSO) time relations between elements of
different layers, for example by measuring the distance to $s$.

On the problematic side, having two elements $v$ and $w$ 
at hand which represent vertices in some layer of the temporal graph,
it seems difficult to get an \MSO-formula which evaluates to true if and only if $v$ and $w$ are generated from the same vertex.
To do so, \citet{MansM14} used an expression $f_V(v) = f_V(w)$.
It is unclear whether $f_V$ is in fact part of the \tps{} or not.
Note that the length of such an expression in terms of the unary relation symbols~$L_v$ depends 
on the number of vertices in the temporal graph.
If the expression $f_V(v) = f_V(w)$ is an short cut for an expression of size at least the number of vertices in the temporal graph,
then Lemmata 13 and 17 and hence Corollaries 14--16 and 18 of \citet{MansM14} break.
We believe that it is rather unlikely that one can provide
arguments to repair the idea of \citet{MansM14}
because of the following example. 

\begin{example}
	\label{exp:npeqp}
	The following is a polynomial-time algorithm for the \NP-complete \prob{3-Coloring} problem on graphs with maximum degree four \cite{DAILEY1980289}:
	Given a graph~$G$ with maximum degree four, construct a temporal graph~$\TG$ with five layers 
on the vertex set~$V(G)$ such that the underlying graph of~$\TG$ is~$G$, and each layer has treewidth one.
Here, the edges in one layer of $\TG$ correspond to one color of a edge-coloring of $G$ with five colors.
Such an edge-coloring can be computed by Vizing's theorem \cite{misra1992constructive}.
Each layer of~$\TG$ and hence the Gaifman graph of the corresponding \tps{} have constant treewidth, because each layer is just a matching.
Thus, 
if there is an \MSO-formula~$X(v,w)$ of constant length which evaluates to true if and only if
two elements $v$ and $w$ are generated from the same vertex, 
then we could easily use this to construct an \MSO-formula of constant length which evaluates to true if and only if 
the underlying graph of $\TG$ and hence $G$ is 3-colorable.
This would imply that~$\classP=\NP$.
\end{example}
In private communication, we discussed our concerns with \citet{MansM14}:
they agreed that one cannot find a constant size \MSO-formula that evaluates to true 
if and only if two elements~$v$ and~$w$ are generated from the same vertex, unless $\classP=\NP$.
\section{Possible Definitions of Temporal Treewidth}
\label{sec:reflections}

\casablanca{Welcome back to the fight. This time I know our side will win.}
Now we embark on the endeavor of finding
useful and interesting definitions for temporal treewidth. To prepare our
journey, we first briefly recapitulate how treewidth (and
other structural graph parameters) have commonly been adapted for the temporal
setting. %
In the majority of cases, structural graph parameters such as treewidth are
transferred to the temporal setting in one of the following two straightforward ways:
\begin{enumerate}
\item Take the maximum over all layer treewidths, resulting in $\ltw$.
\item Take the treewidth of the underlying graph, resulting in $\utw$.
\end{enumerate}
Since the treewidth of a graph does not increase when edges are removed, we
naturally get that for any temporal graph $\TG$ it holds that
$\ltw(\TG)\le\utw(\TG)$.
We can also observe that these two variants of temporal treewidth 
are invariant under reordering of the layers and hence might not be considered
truly temporal since they also apply to the unordered ``multilayer setting''.

There is a further generic way to transfer a structural graph parameter to the temporal setting. 
This one is particularly interesting in the context of problems 
that make use of $\Delta$-time windows\footnote{A \emph{$\Delta$-time window} is a set of $\Delta$ consecutive time steps.}, 
as done in recent work 
on \textsc{Restless Temporal Paths}~\cite{Cas+19}, \textsc{Temporal Clique}~\cite{Ben+19,VLM16,Him+17,MolterNR19}, 
\textsc{Temporal Coloring}~\cite{MMZ19}, 
\textsc{Temporal Matching}~\cite{MertziosMNZZ20}, and 
\textsc{Temporal Vertex Cover}~\cite{Akr+19}. 
In the case of treewidth we call this parameter 
\emph{$\Delta$-slice treewidth}\footnote{To the best of our knowledge, 
the concept of a ``$\Delta$-slice parameter'' was introduced by \citet{Him+17} to define a temporal version of degeneracy. 
It was later also used by \citet{Ben+19}.}, 
and as the name suggests, it depends on 
an additional natural number $\Delta$ that is typically part of the input or the problem specification. %
The $\Delta$-slice treewidth is the maximum of the treewidths of the union graphs of all $\Delta$-time windows,
formally defined as follows:
\begin{definition}[$\Delta$-Slice Treewidth]
For a temporal graph $\TG=(V,E_1,\dots,E_\tau)$ and a natural number $\Delta\le\tau$,
the \emph{$\Delta$-slice treewidth} $\tw_\Delta(\TG)$ of $\TG$ is
defined as 
$$\tw_\Delta(\TG):=\max_{i\in\{1,\ldots,\tau-\Delta+1\}} \tw(G_i^{(\Delta)}),$$
where $G_i^{(\Delta)}=(V,\bigcup_{j\in\{i,\ldots,i+\Delta-1\}} E_j)$.
\end{definition}
It is easy to see the $\Delta$-slice treewidth interpolates between layer
treewidth and underlying treewidth, hence we have that
$\ltw(\TG)\le\tw_\Delta(\TG)\le\utw(\TG)$, for all temporal graphs $\TG$
and all $\Delta\le\tau$.

In light of the known results of temporal graph problems 
where treewidth is used as a parameter (see \cref{tab:results} in \cref{sec:intrac}), 
we observe that even for the largest of our established concepts of treewidth of a temporal graph, 
namely the underlying treewidth, we already obtain para-NP-hardness results for many temporal graph problems. 
Hence, %
temporal treewidth versions 
such as $\Delta$-slice treewidth, 
which are upper-bounded by the underlying treewidth, 
are not desirable since on their own they presumably do not offer new ways to obtain tractability results.

As to islands of tractability %
(see \cref{tab:results} or apply \cref{thm:mso-opt}), 
we find many FPT-algorithms for the combined parameter $\utw+\tau$ 
and for the parameter number $|V|$ of vertices for a variety of temporal graph problems. Further examples of FPT results that are not included in \cref{tab:results} are \textsc{Multistage Vertex Cover}~\cite{FluschnikNRZ19},  \textsc{Restless Temporal Paths}~\cite{Cas+19}, and \textsc{Temporal Coloring}~\cite{MMZ19}.
This means that if we \casablanca{round up the usual suspects} 
of which (combination of) parameters should lower- and upper-bound 
the temporal treewidth in our endeavor of finding useful definitions, 
then we might want to be on the look-out for something between $\utw$ and $\utw+\tau$ or something between~$\utw$ and~$|V|$, 
or something that is incomparable to the aforementioned parameters.

In the following, we are going to discuss three canonical ways to approach defining treewidth for temporal graphs:
\begin{enumerate}
\item Adapting tree decompositions to temporal graphs (\cref{sec:TTDC}).
\item Deriving static graphs from a temporal graph in a natural way and using the treewidth of those graphs (\cref{sec:TSE}).
\item Looking at ways to play cops-and-robber games on temporal graphs (\cref{sec:TCR}).
\end{enumerate}

\subsection{Adaptions of the Tree Decomposition}\label{sec:TTDC}
One beacon of treewidth
applications has always been the tree decomposition. Hence, it is only logical
to begin our quest for temporal treewidth in the decomposition territory.
First, we have to ask ourselves which general properties we want a temporal tree decomposition to have. Should it be temporal as well? We could try to take inspiration from \citet{Bodlaender93a} who showed how to maintain a tree decomposition under edge additions and deletions (when the treewidth is at most two). However, it seems difficult to perform dynamic programming (which is the standard way to design FPT-algorithms for problems parameterized by treewidth) on tree decompositions that keep changing over time. Hence, we focus on \emph{static} tree decompositions for temporal graphs, even though the idea of a temporal tree decomposition that itself is temporal as well probably deserves further consideration.

Second, we have to seek for something to put into our bags. There are two
canonical choices: the vertices $V$ of a temporal graph $\TG=(V,\TE,\tau)$, or its vertex appearances, that is, $V\times\set{\tau}$. If we put the
vertices into our bags, then it seems difficult to end up with something that
is significantly different to the treewidth of the underlying graph and captures
the temporal nature of the setting. So let us see what we can end up with if we
put vertex appearances into the bags. We probably would want to require that
for each temporal edge there is a bag that contains both endpoints of the edge,
which in terms of vertex appearances would be the endpoints of the edge
labeled with the time stamp of the temporal edge. However, if we stop here and
add the straightforward adaptation of the third condition of tree
decompositions, namely that for every vertex appearance, all bags that contain
this vertex appearance should form a connected subtree, then we end up with the
layer treewidth, which is something we do not want. To fix this, we may want to
consider requiring every two vertex appearances with the same vertex and
adjacent time stamps to be contained in at least one bag. This would surely
give us something that is at least as large as the underlying treewidth and at
most as large as the underlying treewidth times the lifetime. The following
definition formalizes this idea.
\begin{definition}[Temporal Tree Decomposition]
 \label{def:temptreedecwidth}
 Let~$\TG=(V,\TE,\tau)$ be a temporal graph.
 A tuple~$\bbT=(T,\{B_u\mid u\in V(T)\})$ consisting of a tree~$T$ and a set of bags~$B_u\subseteq V\times\set{\tau}$ is a \emph{temporal tree decomposition (ttdc)} of~$\TG$ if
 \begin{enumerate}[(i)]
  \item $\bigcup_{u\in V(T)} B_u  = V\times\set{\tau}$,
  \item for every~$(\{v,w\},t)\in \TE$ there is a node~$u\in V(T)$ such that~$(v,t)\in B_u$ and $(w,t)\in B_u$, 
  \item for every $v\in V$ and $t\in\set{\tau-1}$ there is a node~$u\in V(T)$ such that~$(v,t)\in B_u$ and $(v,t+1)\in B_u$, and 
  \item for every~$(v,t)\in V\times\set{\tau}$, the graph~$T[\{u\in V(T)\mid (v,t)\in B_u\}]$ is a tree.
 \end{enumerate}
 The \emph{width} of~$\bbT$ is~$\width(\bbT):=\max_{u\in V(T)}|B_u|-1$.
\end{definition}

As with static treewidth, this definition of a graph decomposition would give a
canonical definition of a temporal treewidth: The temporal treewidth of
a temporal graph~$\TG$ is the minimum width over all temporal tree
decompositions of~$\TG$, that is, \[\ttw(\TG)=\min\limits_{\bbT\text{ is
ttdc of }\TG} \width(\bbT). \] 
As we will see in the next subsection, this definition is
equivalent to using the treewidth of a certain type of static expansion of the
temporal graph. Then it also will become clearer why the proposed definition gives a temporal
treewidth that is at least as large as the underlying treewidth and at most as large
as (roughly) the underlying treewidth times the lifetime.

\subsection{Treewidth of the Static Expansion}\label{sec:TSE} 
Another direction to define a
temporal version of treewidth would be to use the treewidth of static graphs as
we know it, and apply it to a graph that can be naturally derived from a
given temporal graph. The most canonical graph of this type is the \emph{static expansion} 
(see \cref{def:strstaticexp}) 
of a temporal graph which, 
however, 
is typically \emph{directed}\footnote{Note that there are different definitions of static expansion that are typically tailored to the applications they are used in.}. 
One possibility would be to apply
treewidth adaptations for directed graphs~\cite[Chapter~16]{Downey2013OtherWidthMetrics}.
Another possibility is to compute the treewidth of the undirected version of the static expansion of a temporal graph.
Observe that in this case,
we end up with the same temporal treewidth as in \cref{def:temptreedecwidth}.

\begin{observation}\label{obs:staticexp}
Let~$\TG=(V,\TE,\tau)$ be a temporal graph and let $H=(V',A)$ be its static
expansion. Let $G=(V',E)$ with $E=\{\{v,w\}\mid (v,w)\in A\}$ be the
\emph{undirected static expansion} of $\TG$. 
Then 
\[\ttw(\TG)=\tw(G).\]
\end{observation}
We can check that \cref{obs:staticexp} is true by realizing that the bags in a
temporal tree decomposition contain the vertex appearances, which are also the
vertices of a static expansion. Furthermore, the edges of a static expansion
connect all vertex appearances that we want to be together in
at least one bag.

Using this observation, we can also check easily that the claim we made earlier
holds. The precise bounds that we can show are $$\utw(\TG)\le \ttw(\TG)\le
(\utw(\TG)+1)\cdot\tau-1.$$
The lower bound for the temporal treewidth follows from the fact that the underlying graph
is a minor of the undirected static expansion. The upper bound follows from the
observation that the following is a tree decomposition for the undirected static
expansion: take a tree decomposition of the underlying graph and replace every
vertex in every bag by all its appearances. This increases the size of all bags
by a factor of $\tau$.

Now we can also more easily understand how temporal graphs with very small temporal treewidth look like. 
A temporal graph whose temporal treewidth is one necessarily needs to have a forest as underlying graph. 
However, 
even in this case, 
the temporal treewidth can still be as large as $\min\{|V|, \tau\}$ if every edge appears at every time step. 
Take a path as underlying graph as an example where every edge appears at every time step. 
Then the undirected static expansion is a~$|V|\times\tau$-grid. 
In fact, 
as soon as an edge appears at more than one time step, 
the undirected static expansion contains a cycle. 
Hence, 
a temporal graph with temporal treewidth one has a forest as underlying graph and every edge appears in exactly one time step. 
This seems to be a good property of temporal treewidth since many problems are indeed easy to solve on temporal graphs of this form.

\subsection{Playing Cops-and-Robber Games on Temporal Graphs}\label{sec:TCR} 
Since the treewidth of a static graph can be defined via a cops-and-robber game on static graphs (see \cref{sec:TWDef}), 
we can also try to transfer these games to temporal graphs in a meaningful way. 

Recently, \citet{ErlebachS20} investigated a cops-and-robber game on temporal graphs with \emph{infinite} lifetime and periodic edge appearances. Here, whenever the cops and the robber have taken their turn, time moves forward one step, and when making their moves, the cops and the robber can only use edges that are present at the current time.

The first obvious issue with this approach is that the temporal graphs we want to investigate have neither infinite lifetime nor periodic edge appearances. 
If the game would just stop when the lifetime finished and the robber wins if he or she does not get caught, 
then we would need more cops on temporal graphs with shorter lifetime. 
Deriving a temporal treewidth concept from this would lead to the probably undesirable property that temporal graphs with short lifetime have higher treewidth than temporal graphs with a long lifetime. 
To circumvent this, we could repeat the temporal graph ad infinitum. 
This would also make edge appearances periodic. 
However, 
then we will also get the property that the moves a robber can make in this temporal graph is a subset of the moves a robber could make in the underlying graph (or, equivalently, when all edges are always present). 
This means the number of cops necessary to catch a robber in this scenario is upper-bounded by the treewidth of the underlying graph---a property that we do not want to have.

Summarizing, we can say that designing cops-and-robber games on temporal graphs that lead to useful treewidth definitions seems to be a challenging task. 
However, since cops-and-robber games already inherently have a temporal character, 
maybe they are the best-suited way to define temporal treewidth.

\section{Conclusion}

\casablanca{Here's looking at you},
temporal treewidth. 
Indeed, it is a worthwhile endeavor to explore the prospects and limitations of parameters such as 
treewidth transformed to the context of temporal graphs.
A lot of exploration and clarification is yet to do.
So let us agree, in temporal treewidth future we see. Hans, can you?

\paragraph*{Acknowledgments.}
TF acknowledges support by DFG, project TORE (NI~369/18).
HM and MR acknowledge support by DFG, project MATE (NI~369/17).

We thank Mark de Berg, Anne-Sophie Himmel, Frank Kammer, S\'{a}ndor Kisfaludi-Bak, Erik Jan van Leeuwen, and George B.\ Mertzios for their constructive feedback which helped us to improve the presentation of the paper.

We further thank Bernard Mans and Luke Mathieson for helpful discussions concerning the issues presented in \cref{ssec:pitfall}.

\bibliographystyle{abbrvnat}
\bibliography{bodfest}

\appendix

\section{Temporal Graph Problem Zoo}
\label{sec:probzoo}

\decprob{$(\alpha,\beta)$-Temporal Reachability Edge Deletion ($(\alpha,\beta)$-TRED)}
  {A temporal graph~$\TG=(V,\TE,\tau)$ and two integers~$k,h\in\N_0$.}%
  {Is there a subset~$E'\subseteq E(\ug{\TG})$ of the underlying graph's edges with~$|E'|\leq k$ such that in~$\TG-(E' \times \{1,\dots,\tau\})$,
  the size of the set of vertices reachable from every vertex~$s\in V$ via a strict~$(\alpha,\beta)$-temporal path is at most~$h$?} 

\decprob{$(\alpha,\beta)$-Temporal Reachability Time-Edge Deletion ($(\alpha,\beta)$-TRTED)}
  {A temporal graph~$\TG=(V,\TE,\tau)$ and two integers~$k,h\in\N_0$.}%
  {Is there a subset~$\TE'\subseteq \TE$ of temporal edges with~$|\TE'|\leq k$ such that in~$\TG-\TE'$,
  the size of the set of vertices reachable from every vertex~$s\in V$ via a strict~$(\alpha,\beta)$-temporal path is at most~$h$?} 

\decprob{Minimum Single-Source Temporal Connectivity ($r$-MTC)}
  {A temporal graph~$\TG=(V,\TE,\tau)$ with edge weights~$w:\TE\to \Q$, a designated vertex~$r\in V$, and a number~$k\in\Q$.}
  {Is there a temporally $r$-connected spanning subgraph of~$\TG$ of weight at most~$k$?} 

\decprob{Min-Max Reachability Temporal Ordering (Min-Max RTO)}
  {A temporal graph~$\TG=(V,\TE,\tau)$, and an integer~$k\in\N$.}
  {Is there a bijection~$\phi:\set{\tau}\to\set{\tau}$ such that the maximum reachability in~$\TG'=(V,\{(e,\phi(t))\mid (e,t)\in\TE\},\tau)$ 
  is at most~$k$?} 

\decprob{Min Reachability Temporal Merging (MRTM)}
  {A temporal graph~$\TG=(V,\TE,\tau)$, a set of sources $S \subseteq V$, and three integers~$\lambda, \mu, k \in\N$.}
  {Are there $\mu$ disjoint intervals $M_1, \dots, M_\mu \subseteq \set{\tau}$, each of size $\lambda$,
  such that, after merging each of them in $\TG$, the number of vertices reachable from $S$ is at most~$k$?}

 \decprob{Return-To-Base Temporal Graph Exploration (RTB-TGE)}
 {A temporal graph~$\TG=(V,\TE,\tau)$ and a designated vertex~$s\in V$.}
 {Is there a strict temporal walk starting and ending at~$s$ that visits all vertices in~$V$?}
 
\decprob{Temporal Graph Exploration (TGE)}
 {A temporal graph~$\TG=(V,\TE,\tau)$ and a designated vertex~$s\in V$.}
 {Is there a strict temporal walk starting at~$s$ that visits all vertices in~$V$?}

\decprob{Temporal Matching (TM)}
  {A temporal graph~$\TG=(V,\TE,\tau)$ and integers~$k,\Delta\in\N_0$.}
  {Is there a set of~$k$ temporal edges $\TE' \subseteq \TE$
	such that any pair $\{(e, t), (e', t')\} \subseteq \TE'$ has $e \cap e' = \emptyset$ or $\abs{t-t'} \geq \Delta$?} 

\decprob{Temporal Separation (TS)}
  {A temporal graph~$\TG=(V,\TE,\tau)$, two designated vertices~$s,z\in V$, and an integer~$k\in\N$.}
  {Is there a temporal $s$-$z$~separator of size at most~$k$ in~$\TG$?} 

\end{document}